%% file: ms.tex
\documentclass{ar-1col}

\usepackage{natbib}
\usepackage{url}
\urldef\videourl\url{www.dartmouth.edu/~hickox/hickox_alexander_AGN.php}

\input{aasjourn.tex}

\newcommand{\fnt}{\footnotetext}
\newcommand{\fnm}{\footnotemark}

\jname{Annu.\ Rev.\ Astron.\ Astrophys.\ 2018. 56}
\jvol{56}
\jyear{2018}
\doi{10.1146/((please add article doi))}

\begin{document}

\markboth{R.~C.~Hickox \& D.~M.~Alexander}{Obscured AGN}

\title{Obscured Active Galactic Nuclei}

\author{Ryan~C.~Hickox$^1$ and David~M.~Alexander$^2$
\affil{$^1$Department of Physics \& Astronomy, Dartmouth College, Hanover, NH 03755, USA; email: Ryan.C.Hickox@dartmouth.edu}
\affil{$^2$Centre for Extragalactic Astronomy, Department of Physics, Durham University, Durham, DH1 3LE, UK; email: d.m.alexander@durham.ac.uk}}

\begin{abstract}

Active Galactic Nuclei (AGN) are powered by the accretion of material
onto a supermassive black hole (SMBH), and are among the most luminous
objects in the Universe. However, the huge radiative power of most AGN
cannot be seen directly, as the accretion is ``hidden'' behind gas and
dust that absorbs many of the characteristic observational
signatures. This obscuration presents an important challenge for uncovering
the complete AGN population and understanding the cosmic evolution of
SMBHs. In this review we describe a broad range of multi-wavelength
techniques that are currently employed to identify obscured AGN, and
assess the reliability and completeness of each technique. We follow
with a discussion of the demographics of obscured AGN activity,
explore the nature and physical scales of the obscuring material, and
assess the implications of obscured AGN for observational
cosmology. We conclude with an outline of the prospects for future
progress from both observations and theoretical models, and highlight
some of the key outstanding questions.

\end{abstract}

\begin{keywords}
active galaxies, AGN surveys, black holes, central torus, mergers, obscuration
\end{keywords}
\maketitle

\tableofcontents

\section{INTRODUCTION}

\label{sec:intro}

An AGN is the observed manifestation of gas accretion onto a
supermassive black hole (SMBH). In the broadly accepted view of AGN,
the accretion of gas around the SMBH produces an optically thick disk
of material (termed the ``accretion disk''), which emits thermally due
to viscosity within the disk \citep[e.g.,][]{shak73,rees84agn}. The
gas within the accretion disk has a wide range of temperatures (with
the temperature an inverse function of the distance from the SMBH)
and, consequently, the emission is produced over a broad wavelength
range (termed the spectral energy distribution; SED).\footnote{We note
  that this model may only be appropriate for high-accretion rate AGN
  (typical $L/L_{\rm Edd}>10^{-3}$), which are the focus of this
  review. See \citet{done07disk} and \citet{yuan14hot} for discussions
  of low-accretion rate systems.}  For the accretion disk of a typical
AGN, the range of gas temperatures is likely to be
T~$\approx$~$10^4$--$10^5$~K and, therefore, the majority of the
emission from the accretion disk will be at $\approx$~30--300~nm
(i.e.,\ at UV--optical wavelengths).

\begin{figure}[t]
\includegraphics[width=4.5in]{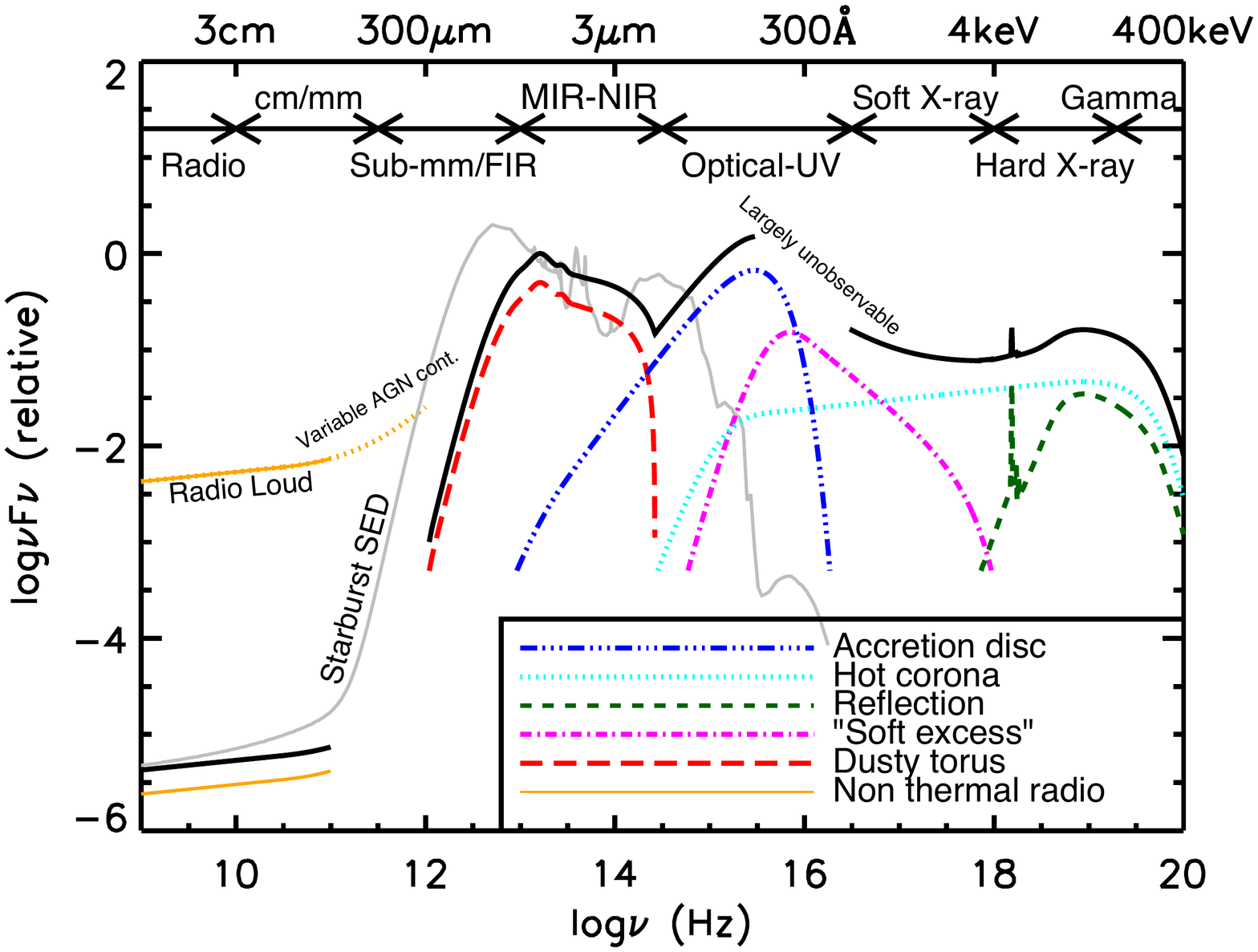}
\caption{Schematic representation of the spectral energy distribution
  (SED) of an unobscured AGN (black curve), separated into the main
  physical components (as indicated by the colored curves) and
  compared to the SED of a star-forming galaxy (light grey
  curve). Figure from \citet[][]{harrison2014}, courtesy of
  C.~M.~Harrison.}
\label{fig:sed}
\end{figure}

The SED of an AGN accretion disk is distinct from that of other
astrophysical sources, making them comparatively easy to identify; see
{\bf Figure \ref{fig:sed}} for the different SEDs between a
star-forming galaxy and the accretion disk. This is fortuitious since
the accretion disk is small and unresolved for even the closest and
brightest AGN (i.e.,\ light hours--light days in physical size). The
accretion of gas onto a SMBH is an exceptionally efficient process
($\approx$~5--42\% of the mass is ultimately converted into emission,
depending on the spin of the SMBH; e.g.,\ \citealt{kerr63, shap83})
and thus large luminosities can be produced for a modest amount
of accretion, allowing for luminous AGN to be detected out to high
redshifts. Indeed, luminous AGN are the most powerful non-explosive
objects in the Universe.

The first systematic studies of AGN occured over 50 years ago and led
to the definition of some of the main classes \citep[e.g.,][]{seyf43,
  baad54radio1, schm63}: Seyfert galaxies, radio galaxies, and
quasars. It is not the objective of this review to describe the
menagerie of different AGN classes, which has been extensively covered
in \citet{pado17agn}. However, we note that in the current parlence,
``Seyfert galaxies'' is often used to indicate AGN of low--moderate
luminosity ($L_{\rm bol}\approx10^{42}$--$10^{45}$~erg~s$^{-1}$) while
``quasars'' is often used to indicate AGN of high luminosity ($L_{\rm
  bol}>10^{45}$~erg~s$^{-1}$). $L_{\rm bol}$ is the bolometric
luminosity of the AGN, which corresponds to the total luminosity
produced (or inferred, for systems where the accretion disk emission
is not directly detected) by the accretion disk.

\begin{figure}[h]
\includegraphics[width=5in]{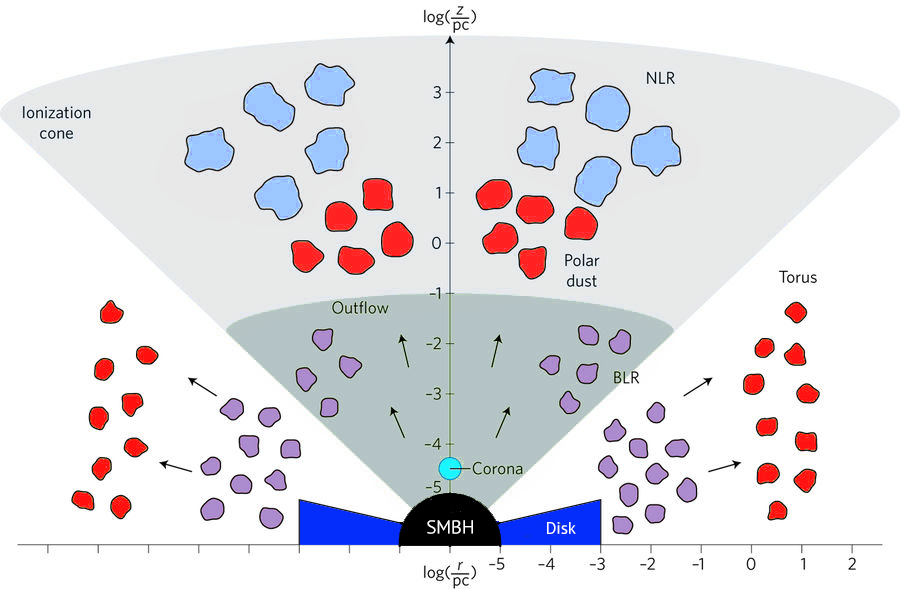}
\caption{Schematic representation of the AGN physical model,
  illustrating the broad scales of the key regions. The accretion
  disk, corona, broad-line region (BLR), and the dusty torus reside
  within the gravitational influence of the SMBH. The disk, corona, and torus (including polar dust clouds) are colored corresponding to the lines showing their contributions to the SED in Figure~\ref{fig:sed}. The narrow-line
  region (NLR) is on a larger scale and under the gravitational
  influence of the host galaxy. Adapted from \citet{ramo17nucl}, courtesy of C.~Ramos Almeida and C.~Ricci.}
\label{fig:structure}
\end{figure}

\subsection{A brief overview of obscured AGN}

\label{sec:overview}

The focus of this review is obscured AGN. These are systems
where the emission from the accretion disk is not directly detected
due to the presence of material between the accretion disk and the
observer. In a general sense, obscuration is defined as anything that
absorbs emission and/or scatters a large fraction away from the line
of sight of the observer. In astrophysical sources the obscuring
medium is typically composed of dust and/or gas. Dust is the common
term used to describe solid-state structures, which are typically
carbonaceous grains and amorphous silicate grains (see
\citealt{drai03dust} for a review). Gas is the term used to describe a
broad range of gaseous states, from fully ionized gas, including
electrons and protons, to neutral gas and molecular compounds. Dust
dominates the source of obscuration at UV--infrared (IR) wavelengths,
while gas dominates the absorption at X-ray energies. However, the
impact of the obscuring material on the detection of the
accretion-disk emission is dependent on wavelength. Physically this is
referred to as the optical depth, which is the product of the opacity
and density of the material ($\kappa_{\lambda}$; $\rho$) and the path
length ($s$); for example, see \citet{rybi79} for a general
overview. A low optical depth indicates that a small fraction of the
emission will be absorbed while a high optical depth implies the
converse.

For many (probably the majority of) obscured AGN, the obscuration
occurs in the close vicinity of the accretion disk and lies within the
gravitational influence of the SMBH. In the favoured picture for the
physical structure of AGN (termed the ``unified model'' of AGN ;
e.g.,\ \citealt{anto93, urry95, netz15unified}), the accretion disk is
surrounded by a geometrically and optically thick dusty and molecular
``torus'' (often referred to as the ``dusty torus''); see {\bf Figure
  2} for a schematic of the AGN physical model from \citet{ramo17nucl}. The torus is expected
to be within the gravitational influence of the SMBH and could be
considered, in a broad sense, the cool outer regions of the accretion
disk where molecules and dust grains can form. The anisotropic nature
of the torus means that for some lines of sight the accretion-disk
emission is directly detected while for others it is obscured by the
dust and gas within the torus. However, the obscuration can also come
from the host galaxy (e.g.,\ from dust-obscured star-forming regions;
dust lanes) and is likely to be more significant for inclined and
edge-on galaxies and for galaxies in gas-rich mergers since, on
average, the typical optical depth along a given line of sight will be
higher than for face-on normal galaxies \citep[e.g.,][]{goul12comp,
  buch17galobsc}.

\begin{figure}[t]
\includegraphics[width=\textwidth]{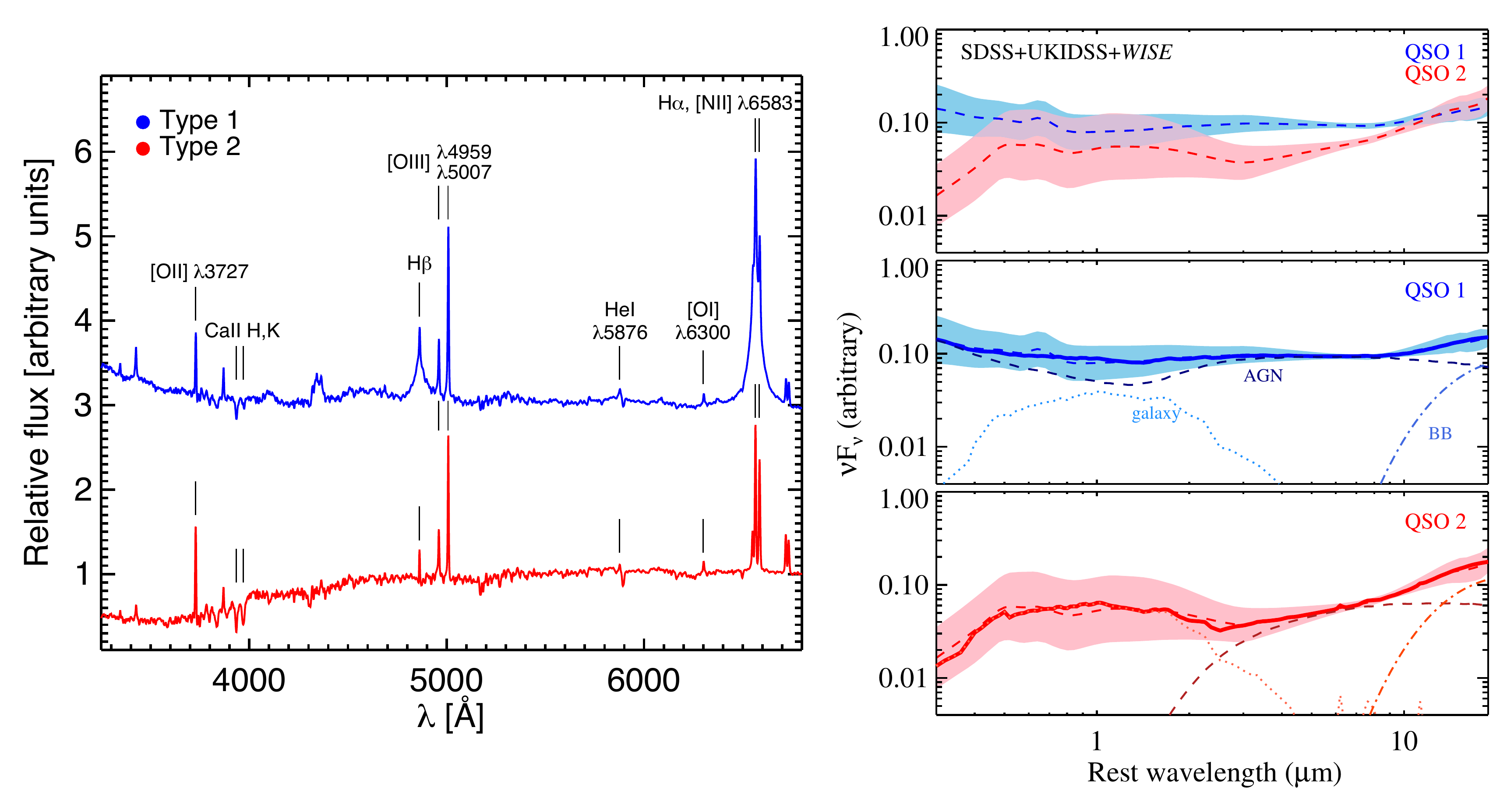}
\caption{(Left) Composite optical spectrum of Type 1 (blue) and Type 2
  (red) AGN from SDSS, adapted from \citet{dipo18outflow}, with the
  prominent emission lines highlighted.  The primary differences
  between the spectra are the presence of bluer nuclear continuum and
  (by definition) broad permitted emission lines in the Type 1 AGN,
  while similar narrow AGN lines are observed in both spectra. (Right)
  Composite optical--MIR SEDs of Type 1 and Type 2 quasars selected
  from SDSS, using data from SDSS, UKIDSS, and {\em WISE}
  \citep{hick17qso}. The composite SEDs are modeled with contributions
  from a (reddened) AGN (dashed lines), the host galaxy (dotted
  lines), and an empirical blackbody component representing emission
  from cooler dust (dot-dashed lines). The SEDs are dramatically
  different in the optical due to reddening of the AGN continuum, but
  are very similar in the mid-IR, highlighting the power of mid-IR
  observations to select obscured AGN. Figures modified from \citet{dipo18outflow}, courtesy of M.~DiPompeo, and \citet{hick17qso}.}
\label{fig:spectrum}
\end{figure}

In addition to the accretion disk and torus there are two other key
regions that we consider in this review for the identification of AGN:
the broad line region (BLR) and the narrow line region (NLR). The BLR
and NLR are defined based on the velocity width of the detected
emission lines in AGN. Empirically, the BLR contains gas with a
broader distribution of velocities than the NLR; the velocity width of
the NLR is often constrained from the forbidden emission lines since
the gas density in the BLR is too high for forbidden transitions. The
typical range of velocity widths for the gas in the BLR is
$\approx$~1,000--10,000~km~s$^{-1}$ while for the NLR it is
$\approx$~100--500~km~s$^{-1}$ \citep[e.g.,][]{pado17agn}. The
different velocity widths of the BLR and the NLR are due to the
relative location of the gas with respect to the SMBH. The gas in the
BLR lies within the gravitational influence of the SMBH and,
consequently, resides close to the accretion disk and is typically
undetected in obscured AGN; see {\bf Figures
  \ref{fig:structure}--\ref{fig:spectrum}}. By comparison, the NLR gas
lies under the gravitational influence of the host galaxy
\citep[e.g.,][]{ho2009} and is produced on larger scales. The bulk of NLR emission generally originates within the central kpc \citep[e.g.,][]{hump15nlr, vill16nlr}, but for some systems emission from gas ionized by the AGN is observed on the scale of the entire galaxy, out to $\sim$10 kpc \citep[e.g.,][]{liu13nlr, hain13salt, hain14nlr}. Due to its extent, the NLR
emission will generally not be obscured by the torus, although a large fraction
of the emission could be obscured by dust in the galaxy.

The classical definition of an obscured AGN is the absence of emission
from the BLR in the optical waveband \citep[e.g.,][]{anto93}. This
corresponds to a typical obscuring screen (or ``extinction'') from
dust of $\approx$~5--10~mags (typically defined in the $V$-band at
550~nm; i.e.,\ $A_{\rm V}=$~5--10~mag;
e.g.,\ \citealt[][]{burt16obsc,schnorr2016}). For typical dust-to-gas
ratios (e.g.,\ as measured in the Galaxy; \citealt[][]{predehl1995})
this corresponds to an equivalent absorbing column density from gas
measured in the X-ray band of $N_{\rm H}>10^{22}$~cm$^{-2}$. The NLR
can be detected in both obscured and unobscured AGN while the BLR is
only expected to be detected in unobscured AGN; see {\bf Figure
  \ref{fig:spectrum}}.

The absence of direct emission from the accretion disk makes obscured
AGN more challenging to identify than unobscured AGN for two key
reasons:

\begin{enumerate}

\item diminished emission: the obscuring material reduces, and in
  extreme cases completely extinguishes, the emission from the AGN; and

\item host-galaxy dilution: the emission from other physical processes
  in the galaxy (e.g.,\ the emission from starlight or star formation)
  dilutes or overwhelms the diminished emission from the AGN, in
  extreme cases making the system indistinguishable from that of a
  galaxy.

\end{enumerate}
  
The impact of these two effects is dependent on the wavelength, the
amount of obscuration, and the relative ratio of the observed emission
from the AGN and the host galaxy. This is illustrated in {\bf Figure
  \ref{fig:sedim}}, which shows schematic multi-wavelength images and SEDs for an
AGN plus the host galaxy for varying levels of nuclear obscuration
(parameterized by the hydrogen column density $N_{\rm H}$) and the strength of the AGN emission
relative to the host galaxy (given by the fraction of the intrinsic
emission from the AGN at 1~$\mu$m, $f_{\rm AGN}$). It is immediately
clear that some AGN signatures are heavily suppressed due to
obscuration or dilution from the host-galaxy emission, while others
remain visible; however the precise observability of these signatures
depends on many parameters including the AGN luminosity and the shape
of its intrinsic spectrum, the characteristics of the host galaxy, and
the geometry and physical nature of the obscuring material. The
wavelength dependent impact of obscuration and host-galaxy dilution on
the identification of obscured AGN are discussed in more detail in
Section \ref{sec:identification}.

\subsection{The importance of identifying obscured AGN}

\label{sec:importance}

The identification of obscured AGN has broad implications for
observational cosmology. The majority of the AGN population is
obscured and therefore the construction of a complete census of AGN
activity requires the identification of both obscured and unobscured
sources. A complete census of AGN activity is required to reliably measure
the cosmological buildup of SMBHs and to place fundamental
constraints on the average radiative efficiency of SMBH growth; we
explore this more in Section \ref{sec:cosmology}. In the unified AGN
model the difference between an obscured and unobscured AGN is the
orientation of the dusty torus with respect to the
observer. Therefore, in some sense the identification of all obscured
AGN may appear to be just a simple book-keeping exercise (i.e.,\ just
accounting for the fraction of the AGN population that are obscured
and therefore not included in unobscured AGN selection approaches). However, the
obscured AGN fraction is found to be a function of AGN luminosity and
potentially redshift, and is therefore not simply a single value (see
Section \ref{sec:demographics}). Furthermore, obscured AGN are more
likely to be found in more dust and gas rich environments than
unobscured AGN and, therefore, the lack of a complete census of
obscured AGN could give a skewed view of the host-galaxy and
larger-scale environments in which AGN reside (see Sections
\ref{sec:physics} and \ref{sec:cosmology}).

This review aims to provide an overview of our observational and theoretical understanding of obscured AGN. In Section \ref{sec:identification} we describe the
challenges in identifying obscured AGN and assess the effectiveness of
the most common techniques, emphasizing two key factors: how reliable
and how complete a given technique is for the identification of
obscured AGN. In Section \ref{sec:demographics} we present the overall
demographics of obscured AGN activity. In Section \ref{sec:physics} we
discuss the physical nature of obscuration in AGN and consider the
relative contributions of obscuration from the nuclear torus, nuclear
starburst regions, and structures on the scale of the host galaxy. In Section
\ref{sec:cosmology} we assess the implications of obscured AGN for
observational cosmology and explore the significance of obscured AGN
for SMBH--galaxy growth and the cosmological radiative efficiency of
SMBH growth. Finally in Section \ref{sec:conclusions} we summarise our
conclusions and discuss the prospects for future observations and
theoretical models to advance the study of obscured AGN. We note that
due to space limitations we have had to be selective in our choice of
cited articles and we apologize in advance for the many significant
studies that we have been been unable to include. We encourage the reader to
also consult the following complementary reviews on the identification
of AGN, the co-evolution of AGN and galaxies, and the unified model of
AGN: \citet{ho08llagn}, \citet{alex12bh}, \citet{heck14bhgal},
\citet{bran15xray}, \citet{netz15unified}, \citet{pado17agn}, and
\citet{ramo17nucl}.

\section{IDENTIFICATION OF OBSCURED AGN}

\begin{figure}[t]
\includegraphics[width=\textwidth]{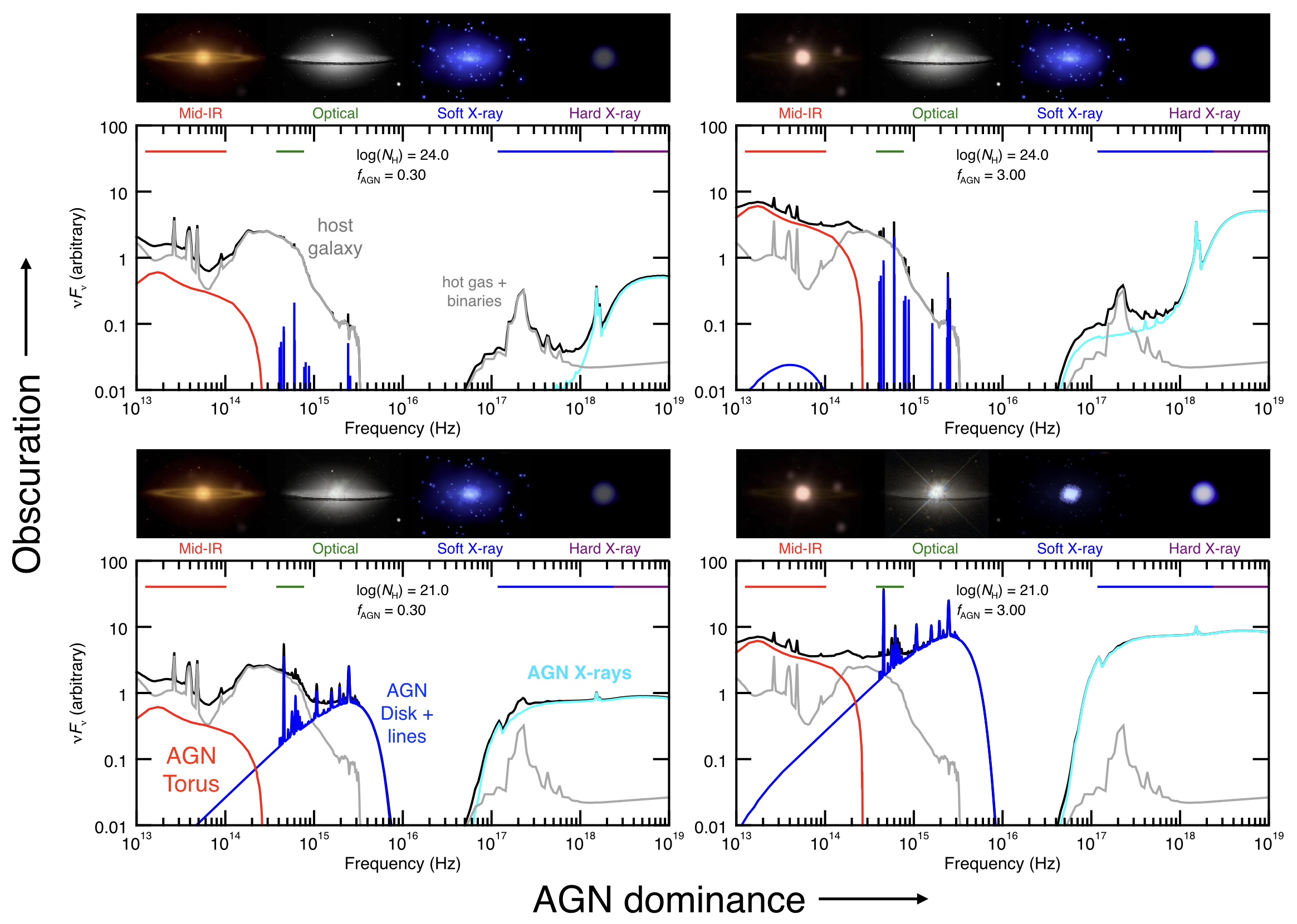}
\caption{Schematic demonstrations of the multi-wavelength
  observational signatures of an AGN as a function of nuclear
  obscuration (increasing from bottom to top) and the relative
  luminosity of the AGN to the host galaxy (increasing from left to
  right). The nuclear obscuration is parameterised by $N_{\rm H}$ and
  the relative luminosity of the AGN to the host galaxy is based on
  the fraction of the intrinsic emission from the AGN at 1~$\mu$m
  ($f_{\rm AGN}$). The bottom component of each panel shows the
  overall model SED (red: AGN component; green: host-galaxy component;
  black: combination of AGN and host galaxy), indicating the emission
  from the host galaxy and the AGN. IR and optical spectral components are taken from \citet{harrison2014} and \citep{asse10agntemp}, optical AGN lines from \citet{vand01}, and X-ray spectra from \citet{revn08xgal} and \citet{balo18borus}. The top component of each panel
  indicates the broad features predicted to be observed in imaging
  data in the mid-IR, optical, and X-ray bands on the basis of the
  model SED, based on images of the galaxy M104 and quasar 3C 273 (images courtesy NASA). {\bf An animation of this figure}, showing the effects of continuous changes in $N_{\rm H}$ and $f_{\rm AGN}$, is available at \videourl.}
\label{fig:sedim}
\end{figure}

\label{sec:identification}

\begin{figure}[t]
\includegraphics[width=\textwidth]{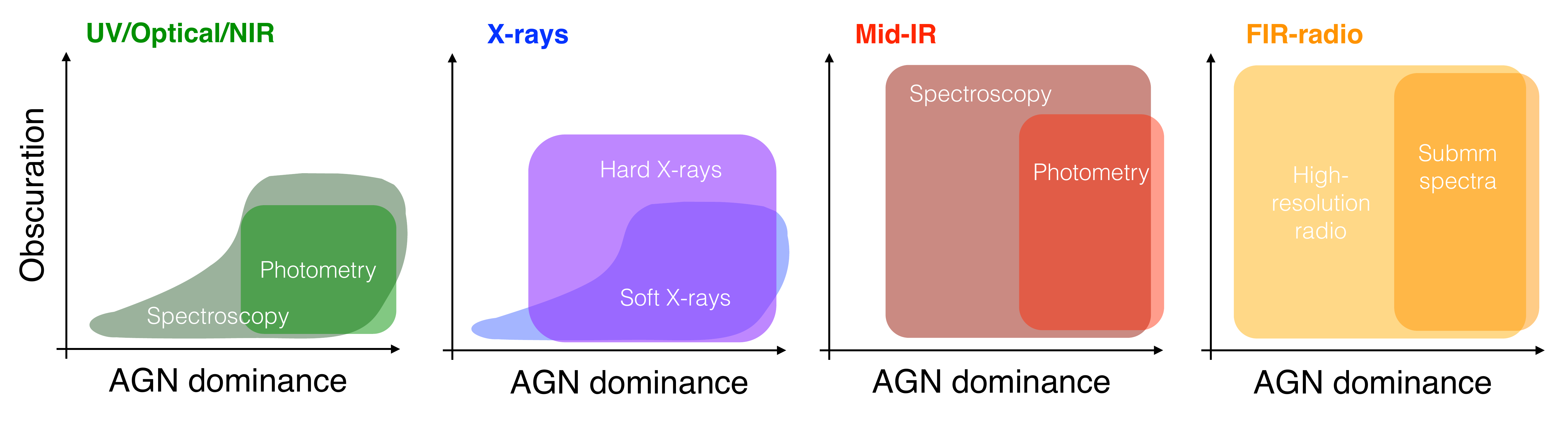}
\caption{Schematic diagram to broadly illustrate the effectiveness of
  different techniques in the identification of AGN for a range in
  nuclear obscuration and the relative luminosity of the AGN to the
  host galaxy, as defined in {\bf Figure \ref{fig:sedim}}. The shaded
  regions indicate where the given technique is expected to be
  effective in identifying AGN activity within the obscuration--AGN
  dominance parameter space; despite the hard edges, these shaded
  regions should only be considered as broadly indicative.}
\label{fig:detschem}
\end{figure}

In this section we give an overview of the variety of methods commonly
used to identify and characterize obscured AGN. We have divided this
section into separate sub sections to correspond to the various
wavebands used to identify obscured AGN: UV--near-IR (0.1--3~$\mu$m),
X-ray, mid-IR (3--30~$\mu$m), and the far-IR--radio ($>30$~$\mu$m); we
do not consider the selection of obscured AGN at gamma-ray wavelengths
since with current technology the majority of the gamma-ray detected
AGN are highly beamed unobscured AGN (see Section 6 of
\citealt{pado17agn} for a recent review). The selection of these
wavebands correspond to those typically adopted in the literature and
are mostly defined by the different technology required to observe in
each waveband; however, we note that each waveband also broadly
corresponds to a specific physical component within the overall AGN
SED (see {\bf Figure \ref{fig:sed}}).

In each sub section we briefly describe the origin of the AGN emission
in that waveband and discuss the impact that obscuration and
host-galaxy dilution has on the identification of AGN activity. We
then describe some of the common techniques adopted to identify AGN
and qualitatively assess two key factors: the ``reliability'' and
``completeness'' of the technique. The ``reliability'' refers to how
reliable a given technique selects an obscured AGN from other
astrophysical source populations (i.e.,\ how much the contamination
from other source populations effect the selection of obscured
AGN). The ``completeness'' refers to how complete the technique is in
selecting obscured AGN (i.e.,\ high completeness means that the
technique is able to select nearly all obscured AGN). These two
quantities are not necessarily correlated. A given technique may be
able to select all obscured AGN but also be unreliable. An extreme example of
this would be to select all galaxies in the Universe: this approach
will select all obscured AGN, and so will have a high completeness;
however, the majority of the selected sources will not host an
obscured AGN and so the technique will have a low reliability. For the
wavebands where both spectroscopy and broad-band photometry are
adopted to select obscured AGN, we will discuss each separately. To
provide some guidance in advance of our discussion, in {\bf Figure \ref{fig:sedim}} we illustrate the effects of obscuration on the broad-band AGN SED, and in {\bf Figure
  \ref{fig:detschem}} we schematically illustrate the impact that
obscuration and host-galaxy dilution can have on the identification of
AGN for a range of different techniques.

In our discussion we also indicate how accurately the amount of
obscuration can be measured from a given technique. The best methods
for identifying obscured AGN will typically not provide the most
accurate measurements on the amount of obscuration because, by
definition, they are relatively insensitive to the presence of
obscuration (i.e.,\ the optical depth is low and therefore the
signatures of obscuration will not be strong).

We begin our discussion with UV--near-IR selection techniques, which
were the first to identify obscured AGN, followed by a discussion of
identification techniques at X-ray, mid-IR, and the far-IR--radio
wavebands, highlighting the reliability and completeness of the
various techniques for the selection of obscured AGN. This discussion
builds on the recent review by \citet{pado17agn}, with a focus on
obscured AGN. At the end of this section we discuss how a combination
of techniques can be utilized to identify a more complete census of
obscured AGN.

\begin{textbox}[t]\section{COMMON UV/OPTICAL/NEAR-IR SELECTION CRITERIA FOR OBSCURED AGN}

Commonly used criteria for identifying AGN in this waveband include:

\begin{itemize}
\item High ratio of high-excitation to low-excitation emission lines; \item Detection of very high-excitation emission lines (e.g., [NeV]); and \item UV, optical and/or NIR colors characteristic of an AGN accretion disk.
\end{itemize}

Once AGN have been identified, common criteria for classifying the sources as obscured include: 

\begin{itemize}
\item Width of permitted emission lines $<1000$ km s$^{-1}$;
\item High nuclear extinction from spectral analysis or multiwavelength SED fitting; a typical criterion is $A_V > 5$ mags; and \item Weak UV/optical/NIR emission compared to AGN luminosity identified in other wavebands (e.g., X-ray, mid-IR).
\end{itemize}

\end{textbox}

\subsection{Selection of obscured AGN in the ultra-violet to near-infrared waveband}

\label{sec:optical}

Some of the most well-developed selection techniques for obscured AGN are in the UV, optical, and near-IR wavebands. A summary of methods for identifying obscured AGN in these bands is given in the sidebar on ``Common UV/optical/near-IR selection criteria for obscured AGN'', and these are discussed in detail below. 

\subsubsection{Broad-band continuum techniques}

Unobscured AGN are efficiently selected using UV--optical photometry
since the emission from the accretion disk is bright in this waveband
\citep[e.g.,][]{rich01qsocol,pado17agn}; see {\bf Figure
  \ref{fig:sed}}. However, UV--optical photometry is ineffective at
identifying obscured AGN because (1) the optical depth due to dust is
high at UV--optical wavelengths and hence the emission from the
accretion disk is easily obscured (the optical depth increases towards
shorter wavelengths; e.g.,\ \citealt{calz94, drai03dust}) and (2) the
host galaxy is bright at UV--optical wavelengths due to the emission
from stars, which dilutes the weak emission from the obscured AGN. As
a consequence both the reliability and completeness of obscured AGN
selection using UV--optical photometry is low. 

Improvements in the
selection of obscured AGN can be made by extending out to near-IR
wavelengths as the optical depth is substantially lower than at
UV--optical wavelengths. However, the stellar emission from galaxies
typically peaks at near-IR wavelengths, offsetting part of the optical
depth benefit. Consequently, the near-IR photometric selection of AGN
is most effective for luminous AGN with modest amounts of obscuration,
such as dust-reddened quasars where some of the accretion disk and
broad-line emission is visible \citep[e.g.,][]{webs95, glik07redqso};
see Section \ref{sec:evolution} for further discussion. 

The weakness of the AGN continuum of obscured AGN with respect to the host galaxy
at UV--near-IR wavelengths does, however, make this waveband ideal for
studying the host galaxies of obscured AGN \citep[e.g.,][]{kauf03host,
  hick09corr, scha10morph, heck14bhgal}. Correspondingly, UV--optical emission that is faint and/or dominated by the host galaxy can be essential for classifying AGN as obscured when they are identified in other wavebands, as well as measuring the level of obscuration through fitting of the multiwavelength SED using empirical and/or theoretical models with varying levels of obscuration on the AGN component (e.g., \citealt{hick07abs, merl14agnobs}; see {\bf Figure~\ref{fig:spectrum}}).

\subsubsection{Spectroscopic techniques}

\label{sec:optspec}

While UV--near-IR photometry is inefficient at identifying obscured
AGN, UV--near-IR spectroscopy and spectropolarimetry have been
essential tools in the identification and characterisation of obscured
AGN. Optical spectroscopy led to the discovery of Seyfert galaxies
\citep{seyf43} and the identification of the two main spectral classes
of AGN \citep[e.g.,][]{khac74, weed77}: Type 1 systems, where the
bright optical continuum and both broad and narrow emission lines are
observed (i.e.,\ unobscured AGN), and Type 2 systems, where the
optical continuum is weak and only narrow emission lines are observed
(i.e.,\ obscured AGN), as illustrated by the composite spectra shown in {\bf Figure~\ref{fig:spectrum}}. Optical spectropolarimety of Seyfert 2 galaxies
has furthermore shown that many have the features expected for a
Seyfert 1 galaxy (broad emission lines and a strong UV--optical
continuum) when observed in polarized light
\citep[e.g.,][]{anto85ngc1068, tran03unified, mora07ngc2110, ramo16specpol}; this is commonly
referred to as a ``hidden Seyfert 1'', a ``hidden Type 1 AGN'', or a
``hidden BLR''. The Seyfert 1 features are undetected in optical
spectroscopy due to the obscuring torus but are seen in
spectropolarimetry due to the emission being scattered (and hence
polarized), by electrons/dust grains within the NLR, into the line of
sight of the observer. Spectropolarimetry was central to the
development of the unified AGN model and our current picture of the
AGN physical structure.

\begin{figure}[t]
\includegraphics[width=5in]{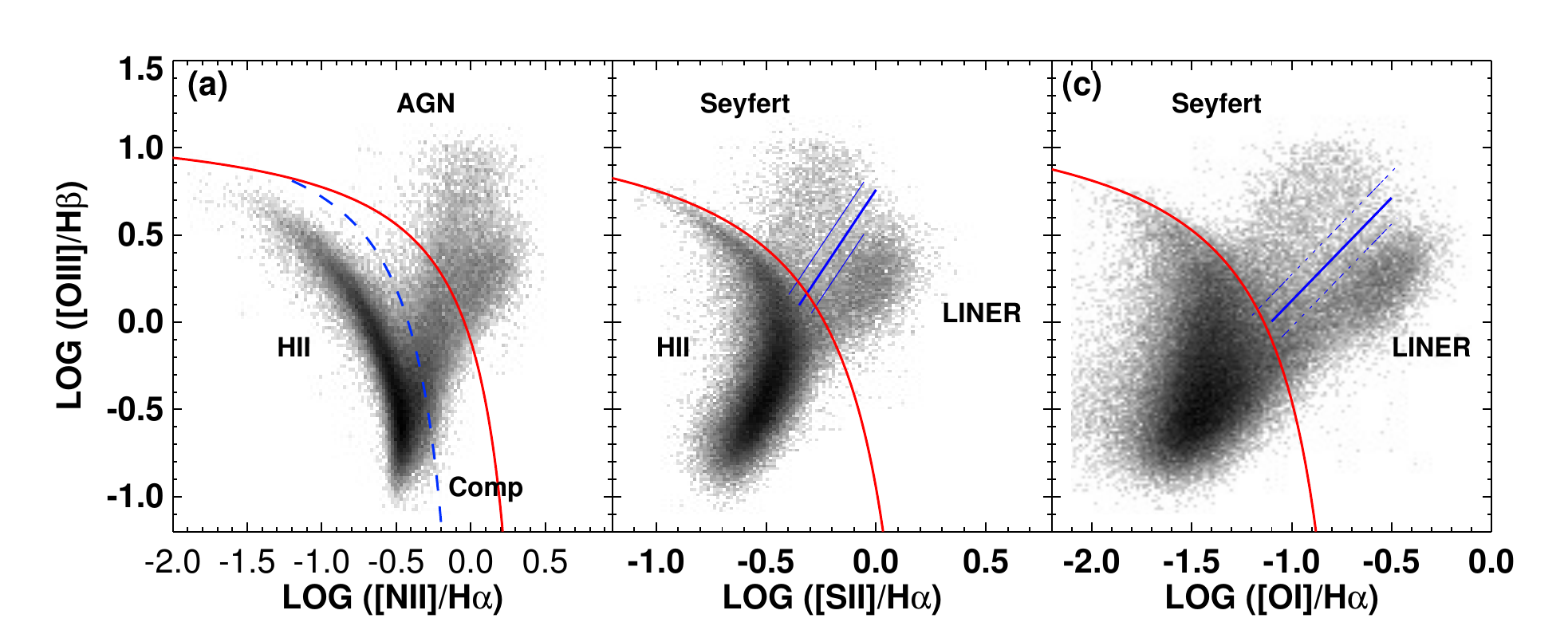}
\caption{Optical emission-line diagnostic diagram to distinguish
  between AGN, star-forming galaxies (H~II), composite AGN--star
  forming systems, and low ionisation nuclear emission region sources
  (LINERs). Figure from \citet{kewl06agn}, courtesy of
  L.~J.~Kewley.}
\label{fig:bpt}
\end{figure}

A key reason why UV--near-IR spectroscopy has been so instrumental in
the identification and characterization of obscured AGN is because the
UV--near-IR waveband is rich in emission lines; some of the prominent optical lines are shown in {\bf Figure
  \ref{fig:spectrum}}. From a combination of emission-line strengths
it is possible to make sensitive measurements of the conditions of the
different gas phases (e.g.,\ from the excitation and ionization
energies and critical densities), allowing for the construction of
powerful emission-line diagnostic diagrams to distinguish obscured AGN
from other astrophysical sources \citep[e.g.,][]{bald81bpt,
  veil87line, ho97agnhost, kewl06agn}. An example suite of
emission-line diagnostic diagrams is shown in {\bf Figure
  \ref{fig:bpt}}. An AGN is distinguished from that of a star-forming
region (i.e,\ HII region) and a LINER (low-ionization nuclear emission
region sources, some of which host low-luminosity AGN;
e.g.,\ \citealt{heck80liner, heck14bhgal}) on the basis of the ratio
of the ionized forbidden line flux to the neutral permitted line flux:
large ratios imply a ``hotter'' (i.e.,\ shorter wavelength and higher
energy) radiation field and betray the presence of the AGN. The
emission-line ratios will be reduced if the AGN resides in a strong
star-forming galaxy until, in extreme cases, the AGN signature is no
longer distinguishable from that of star-forming galaxy. The emission
lines are chosen to have similar wavelengths to reduce the effect of
dust reddening on the emission-line ratio; however, the presence of
obscuration will reduce the strength of the individual emission lines
and could completely extinguish the emission-line
signatures. For applications in which the available emission lines are limited due to the wavelength range of the spectra or the redshift of the source, complementary selection criteria have been developed that use a combination of line ratios and host galaxy properties such as color and mass \citep[e.g.,][]{trou11tbt,yan11aegis, june11agn}. Other studies have identified AGN by the presence of a single very high excitation optical line (commonly [NeV]$\lambda$3426) that is not easily excited by stellar processes \citep[e.g.,][]{gill10nev, mign13nev, verg18nev}.

Emission-line diagnostics provide a reliable
method to identify AGN, although the demarcation curves between AGN
and other source types depend on the metallicity of the systems and
are expected to change with redshift
\citep[e.g.,][]{kewl13bpt1,juneau14}. They also provide a relatively
complete selection of AGN and are able to identify systems with
extreme nuclear obscuration since the NLR emission will not be heavily
extinguished by obscuration of the nucleus on small scales \citep[e.g.,][]{zaka03, lans15nustarqso, hvid18salt}. However, the optical AGN
signatures can be extinguished by dust in the host galaxy or rendered
unidentifiable in systems where the host galaxy emission lines are
significantly brighter than those of the AGN; the former limitation
can be mitigated using spectroscopy at longer wavelengths (e.g.,\ at
mid-IR wavelengths), where the optical depth is lower; see Section
\ref{sec:midir}.

The identification of a Type 2 AGN does not always indicate the
presence of obscuration. At least a fraction of the Type 2 AGN
population appear to be unobscured and intrinsically lack a BLR
\citep[e.g.,][]{pane02seyf2,bian12seyf2}. These systems often appear
to be low-luminosity AGN and it is possible that the accretion rate is
insufficient to allow for the formation of the optically thick
accretion disk \citep[e.g.,][]{elit09blr, trum11agn}; however, some high
accretion rate systems also appear to intrinsically lack a BLR
\citep[e.g.,][]{ho2012,miniutti2013,elitzur2016}, challenging the
simplest versions of the unified AGN model. The optical spectral
properties of some AGN are also found to change from a Type 1 AGN to a
Type 1.8--1.9 AGN (i.e.,\ a system with with most of the features of a
Type 2 AGN but with a weak broad-line component;
\citealt{osterbrock81}); these systems are often called ``changing-look'' AGN. For a fraction of these sources the changing optical
spectral type may be related to changes in the obscuration properties
(see Section \ref{sec:xray} for X-ray evidence of time-variable
absorption in AGN). However, in the majority of cases, it appears to
be related to a decrease in the luminosity of the accretion disk
\citep[e.g.,][]{denn14mrk590, lama15qso, macl16qso}.

UV--near-IR spectroscopy can provide reliable measurements of the
amount of obscuration towards either the BLR or the NLR through the
relative strength of related emission lines. The common adopted
technique is called the ``Balmer decrement'' and is based on measuring
the strength of the Hydrogen emission corresponding to different
excitation states. For example, the H$\alpha$/H$\beta$ flux ratio,
which corresponds to the $n=2$ (i.e.,\ the Balmer series) electron
transitions, will have a fixed ratio for any given conditions of the
gas (e.g.,\ the density, metallicity, temperature;
\citealt{bake38balmer, broc71}). Therefore, since optical depth is
wavelength dependent, the presence of obscuring material will
preferentially affect the H$\beta$ emission (at 486~nm) more strongly
than the H$\alpha$ emission (at 656~nm), and therefore the deviation
of the H$\alpha$/H$\beta$ flux ratio from the expected flux ratio will
indicate the amount of obscuration towards the line emitting regions
\citep[e.g.,][]{ward87,gask17blr}. The Balmer decrement is based on
extinction measurements made from optical emission lines and is
sensitive to modest amounts of obscuration; however, larger amounts of
obscuration can be measured using longer-wavelength emission lines
(e.g.,\ the $n=3$ Paschen series, which are produced in the near-IR
waveband).

\subsection{Selection of obscured AGN in the X-ray waveband}

\label{sec:xray}

X-ray observations provide one of the most reliable and complete
methods for selecting obscured AGN.  A summary of X-ray methods for identifying obscured AGN is given in the sidebar on ``Common X-ray selection criteria for obscured AGN'', and we discuss them in detail below. The X-ray band is defined here as
the energy range of 0.2--200~keV; energies below this range correspond
to UV wavelengths and energies above this range correspond to gamma
rays. The Earth's atmosphere is opaque at X-ray energies and therefore
all sensitive X-ray observations of the cosmos have been obtained from
space (see \citealt{giac09xray} for a review of the history of X-ray
astronomy). It is common practice to divide the X-ray energy band into
the ``soft band'' and ``hard band'', which corresponds broadly to the
sensitivity of the X-ray observatory to the absorbing column density
of gas. Although the definition of these bands can vary from study to
study, in this review we will define the soft band as $<10$~keV and
the hard band as $>10$~keV; on the basis of this definition, some of
the current soft-band observatories are {\it Chandra} and {\it
  XMM-Newton} while some of the current hard-band observatories are
{\it Integral}, {\it NuSTAR}, and {\it Swift}-BAT. The majority of
X-ray observatories have good energy resolution and low background,
allowing for simultaneous photometric and spectroscopic measurements
from individual observations. Hence in this sub section we do not make
a strong distinction between the photometric and spectroscopic
identification of AGN.

\begin{textbox}[t]\section{COMMON X-RAY SELECTION CRITERIA FOR OBSCURED AGN}

Commonly used criteria for identifying AGN in this waveband include: 

\begin{itemize}
\item Observed or intrinsic X-ray luminosity higher than expected for stellar processes (hot gas and X-ray binaries) in the galaxy. A typical criterion is soft (0.5--10 keV) $L_X > 10^{42}$~erg~s$^{-1}$, which is sufficient for all but the most extreme host galaxies; and 
\item Identification of an X-ray point source in high-resolution imaging of the nucleus of the host galaxy (for nearby galaxies, although note caveats in Section \ref{sec:xray}).
\end{itemize}

Once AGN have been identified, common criteria for classifying the sources as obscured include:  

\begin{itemize}
\item X-ray spectral fitting results implying $N_{\rm H} > 10^{22}$ cm$^{-2}$, or equivalent measurements using X-ray hardness ratios; \item Low ratio of observed X-ray luminosity to intrinsic AGN luminosity (usually determined from IR or optical data); and
\item High equivalent width of Fe K$\alpha$ line.
\end{itemize}

\end{textbox}

The X-ray emission from AGN appears to be (near) ubiqitious and is
directly associated with the accretion disk. In unobscured AGN there
is a remarkably tight relationship between the X-ray emission and the
UV--optical emission \citep[e.g.,][]{stef06alphaox, luss16alphaox}.  The
X-ray emission is thought to arise in a ``corona'' above the accretion
disk and is predominantly produced by the inverse Compton scattering
of photons from the accretion disk; however, the lowest energy X-ray
photons can be produced in the inner, and therefore hottest, regions
of the accretion disk. The X-ray emission is then modified from
interactions with material in the accretion disk (and potentially the
host galaxy), such as photoelectric absorption, reflection, and
scattering (see {\bf Figure \ref{fig:sed}}).

The impact of obscuration in the X-ray band is a function of
rest-frame energy, with lower-energy X-ray photons more easily
absorbed than higher-energy X-ray photons (i.e.,\ the optical depth
increases with decreasing energy; \citealt{wilm00abs}). As a
consequence, observatories with sensitivity in the hard band are able
to detect more heavily obscured AGN than observatories with
sensitivity in the soft band; however, we note that it depends on the
redshift of the source since the probed rest-frame energy increases
with redshift in a given energy band. The low optical depth at X-ray
energies, particularly in the hard band, means that the completeness of
obscured AGN selection is high in the X-ray waveband. For example, at X-ray
energies of $>10$~keV significant suppression of the X-ray emission
only occurs at Compton-thick column densities ($N_{\rm
  H}>1.5\times10^{24}$~cm$^{-2}$) due to Compton recoil and subsequent
absorption of the X-ray photons (see \citealt{coma04cthick} for a
review). 

X-ray observations also provide one of the most reliable methods to
identify AGN because the X-ray emission from other astrophysical
processes is typically weak by comparison. The dominant physical
processes for the production of X-ray emission in the host galaxy are
accreting neutron stars and stellar-mass black holes (commonly referred to as X-ray
binaries; see \citealt{fabb06xrb} for a review) and hot gas
($T>10^6$~K). The populations of X-ray binaries are classified into
low-mass X-ray binaries and high-mass X-ray binaries, depending on the
mass of the stellar companion in the binary system, and their integrated
X-ray luminosities are closely related to the mass and star-formation
rate of the galaxy, respectively \citep[e.g.,][]{lehm16galx}. Only the
most massive and strongly star-forming galaxies will produce X-ray
emission of $L_{\rm X}>10^{42}$~erg~s$^{-1}$ at 2--10~keV
\citep[e.g.,][]{alexander2005, wang2013} and the majority of galaxies
will be more than an order of magnitude less luminous in the X-ray band
\citep[e.g.,][]{lehm10sfx, mine12sfx}. Therefore, contamination from
host-galaxy processes is only likely to be an issue for low luminosity
or heavily obscured AGN, where the X-ray emission is suppressed due to
the presence of absorption. For nearby systems, such weak or obscured AGN may be identified using high-resolution X-ray observations if an X-ray point source can reliably associated with the galactic nucleus \citep[e.g.,][]{gall10amuse, she17xray}. However, we caution that there can still be a non-negligible chance that a nuclear X-ray source is an X-ray binary rather than an AGN, highlighting the importance of additional discriminating criteria such as those described in this review.

Since the integrated emission from X-ray
binaries is predominantly observed at $<10$~keV, low luminosity or
heavily obscured AGN can be more reliably identified at $>10$~keV. The
emission from hot gas, either from the host galaxy or from a galaxy
cluster, can be substantial in some sources (up-to
$\approx10^{41}$--$10^{42}$~erg~s$^{-1}$ for AGN in luminous host
galaxies and up-to $\approx10^{44}$--$10^{45}$~erg~s$^{-1}$ for AGN that
reside at the core of a massive galaxy cluster); however, as for the
X-ray binaries, the X-ray emission from the hot gas is predominantly at
low energies ($<$~2--5~keV). Overall, the combination of the low optical
depth and the low contrast between AGN and other astrophysical source
populations at X-ray energies, particularly in the hard band, make X-ray
observations one of the most reliable and complete methods to select
obscured AGN \citep[][]{bran15xray}.

\begin{figure}[t]
\begin{minipage}{2.8in}
\includegraphics[width=2.8in]{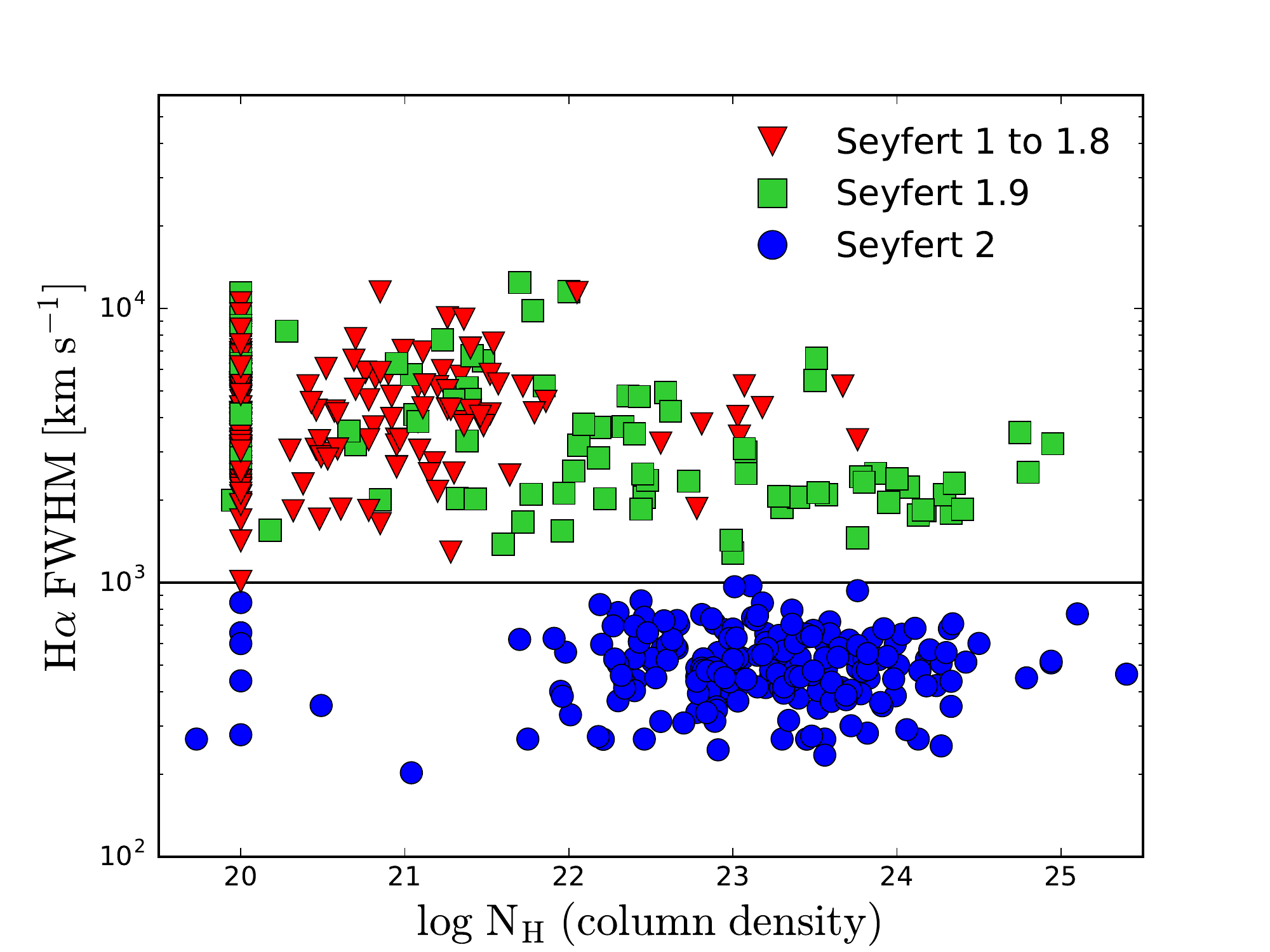}
\end{minipage}
\begin{minipage}{2.2in}
\includegraphics[width=2.2in]{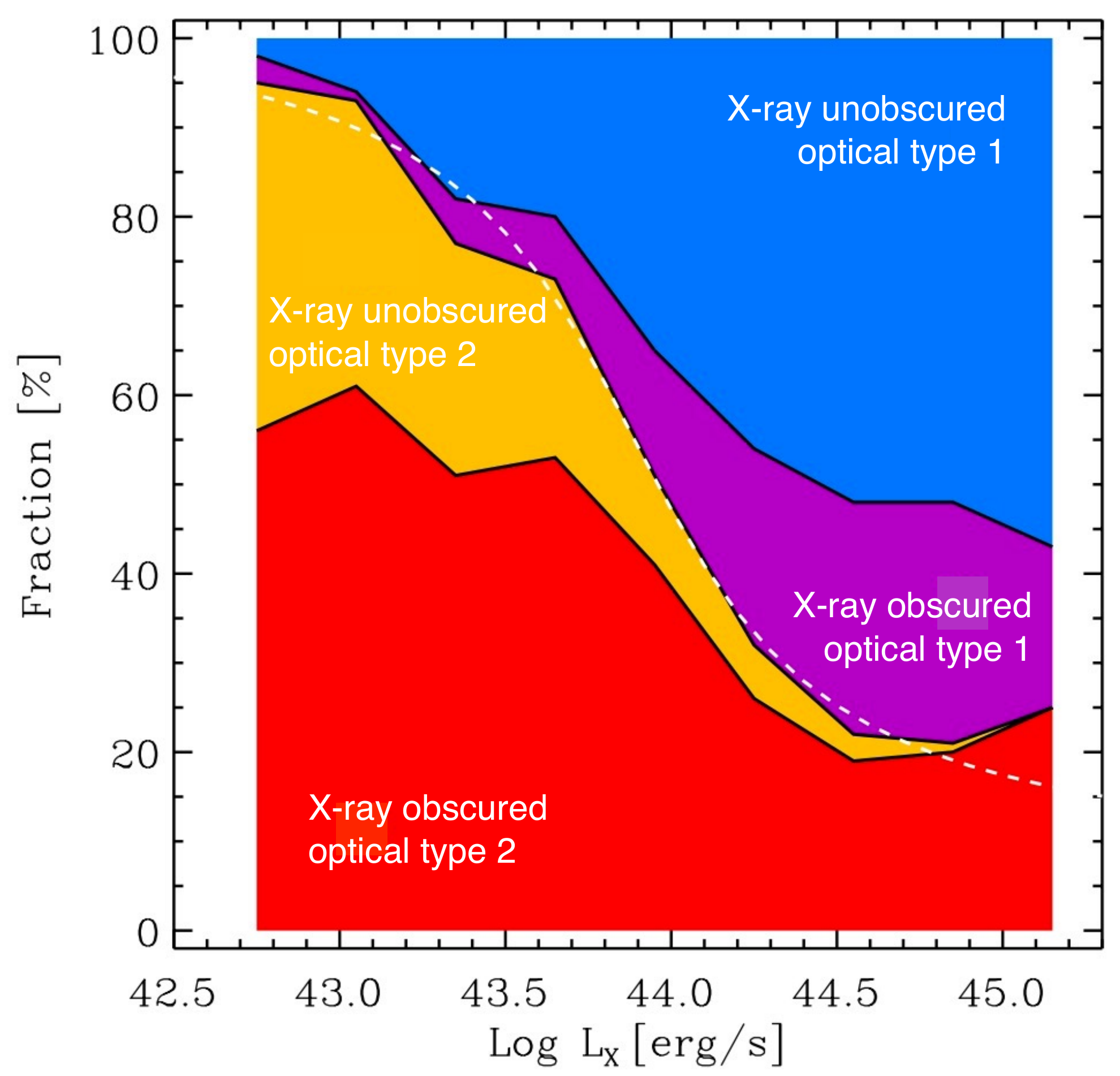}
\end{minipage}
\caption{(Left) $H\alpha$ emission-line width (parameterised as the
  full width half maximum) vs absorbing column densities for local
  X-ray AGN detected by {\it Swift}-BAT. The symbols indicate the
  different optical spectral types of the X-ray AGN and the solid
  horizontal line indicates the broad separation between Type 1 and
  Type 2 AGN in this study. Figure from \citet{koss17bass}, courtesy
  of M.~Koss. (Right) Relative fractions of AGN vs X-ray luminosity
  for the broad redshift range of $z=$~0.3--3.5. The relative
  fractions are distinguished according to their X-ray and optical
  classifications. Figure modified from \citet{merl14agnobs}, courtesy
  of A.~Merloni.}
\label{fig:fobsc}
\end{figure}

The distinctive signatures of absorption in the X-ray spectra of AGN
(illustrated in {\bf Figure~\ref{fig:sedim}}) mean that X-ray spectroscopy provides one of the most accurate methods of identifying obscuration in AGN and measuring
the amount of absorption, at least up to Compton-thin
absorbing column densities \citep[e.g.,][]{done10xray}. For sources with relatively few detected X-ray counts, the level of absorption can be approximated using the hardness ratio (the ratio of source counts in different X-ray bands; \citealt{park06}) and source redshift \citep[e.g.,][]{merl14agnobs}.

The signatures of Compton-thick absorption are more challenging to detect, but can be identified in high signal-to-noise ratio X-ray spectra. The identification
of a strong reflection component at $E>10$ keV and a prominent Fe~K$\alpha$ emission
line at 6.4 keV (with equivalent width typically $>$1 keV) are clear signatures of Compton-thick absorption \citep[e.g.,][]{mush93, leve06comp, done10xray}. The presence of Compton-thick
material can also be inferred through the comparison of the observed X-ray
luminosity to a proxy (such as mid-IR or optical narrow-line emission) for the intrinsic  X-ray luminosity,  although this is a less reliable
approach; see Section \ref{sec:multiwave}. On the basis of current
constraints, the fraction of the AGN population that is Compton thick
is high ($\approx$~30--50\%; e.g.,\ \citealt{burl11batagn,
  ricc15cthick}); see Section \ref{sec:demographics}.

The source of the obscuration at X-ray energies (gas) is different to
the source of the obscuration at longer wavelengths (dust) and
may occur on different physical size scales. Despite this, good
agreement is found between the X-ray signatures of absorption and the
optical spectral types of AGN if $N_{\rm H}=10^{22}$~cm$^{-2}$ is
taken as the threshold between X-ray absorbed and X-ray unabsorbed AGN
($\approx$~80--90\% agreement in the X-ray and optical obscuration
signatures; e.g.,\ \citealt{mali12integral, merl14agnobs, burt16obsc, koss17bass});
see {\bf Figure \ref{fig:fobsc}}. 

The X-ray absorption in the majority
of AGN is not constant but is seen to vary on time scales of days to years
\citep[e.g.,][]{risa02agnnh,bianchi2009}, indicating that the
absorbing gas is distributed in compact clouds; individual AGN have
even been seen to change from Compton thin to Compton thick levels of
absorption and vice versa (e.g.,\ \citealt{matt03xagn,
  risa05ngc1365}). Current constraints suggest that much of the variable
absorption of the X-ray emission occurs in the BLR rather than in the
dusty torus.

\begin{textbox}[t]\section{COMMON MID-IR SELECTION CRITERIA FOR OBSCURED AGN}

Commonly used criteria for identifying AGN in this waveband include:

\begin{itemize}
\item Color diagnostics from mid-IR photometry; \item Significant contribution of AGN to mid-IR emission, from measurement of features in the mid-IR spectrum or fitting of AGN and galaxy templates to the mid-IR SED;
\item Detection of very high-excitation emission lines (i.e. [NeV], [Ne VI]); and
\item Identification of point source in high-resolution observations of galactic nucleus (for nearby galaxies).
\end{itemize}

Once AGN have been identified, common criteria for classifying the sources as obscured include:  

\begin{itemize}
\item Red UV/optical--mid-IR photometric colors; \item High nuclear extinction (for example, $A_V > 5$ mags) from spectral analysis or optical/IR SED fitting; and \item Detection of solid-state absorption features in mid-IR spectrum (particularly the Si features at 9.7 and 18 $\mu$m).
\end{itemize}

\end{textbox}

\subsection{Selection of obscured AGN in the mid-infrared waveband}

\label{sec:midir}

The mid-IR waveband provides an efficient and effective selection of
obscured AGN. A summary of mid-IR methods for identifying obscured AGN is given in the sidebar on ``Common mid-IR selection criteria for obscured AGN'', and we discuss them in detail below. 
The mid-IR waveband is defined here as 3--30~$\mu$m, which
broadly corresponds to the infrared wavelengths where the emission from
the AGN is most distinct from that of the host galaxy; see {\bf Figure
\ref{fig:sed}}. Due to strong absorption in the terrestrial atmosphere,
the majority of the mid-IR waveband can only be efficiently observed
from high in the atmosphere and, ideally, from space; some of the
observatories with sensitivity at mid-IR wavelengths, from the past and
present, are {\em Akari}, {\em IRAS}, {\em ISO}, {\em SOFIA}, {\em
Spitzer}, and {\em WISE} \citep[see][for a review]{lutz14ir}. There are
several narrow wavebands where the mid-IR emission can penetrate through
the atmosphere and be detected using ground-based telescopes. However,
the sensitivity of ground-based telescopes to the detection of mid-IR
emission is low when compared to those from space since the mid-IR
waveband corresponds to thermal emission at $\approx$~100--1000~K and
therefore any ``hot'' objects (including astronomers!) are a significant
source of background noise; by comparison, space-based observatories can
be efficiently cooled, reducing the thermal background noise. However,
ground-based telescopes often have significantly larger mirrors than the
space-based telescopes, allowing for higher-resolution imaging.

\subsubsection{Broad-band continuum techniques}

\label{sec:mircont}

AGN are bright in the mid-IR waveband due to the thermal emission from
warm--hot dust in the torus, which is heated by the absorption of
shorter-wavelength photons from the accretion disk. The strength of
the mid-IR emission from the AGN depends on the covering factor of the
dust around the accretion disk (i.e.,\ the fraction of the photons
from the accretion disk that are absorbed by the dust); see Section
\ref{sec:torus}. The optical depth is low at mid-IR wavelengths and
therefore, unlike the UV--near-IR waveband, the emission is not
strongly suppressed by the obscuring dust. Consequently, the
completeness of the selection of obscured AGN at mid-IR wavelengths
can be high, although see some of the caveats noted below. However,
other astrophysical sources also produce strong infrared emission,
most notably dust-obscured star formation from the host galaxy, which
effects the overall reliability of AGN selection in the mid-IR
waveband, particularly for lower-luminosity systems where the emission
from the AGN can be weak compared to that of the host
galaxy. Fortunately, the SEDs of AGN and star-forming galaxies
significantly differ (e.g.,\ star-forming galaxies have ``cooler''
SEDs; see {\bf Figure \ref{fig:sed}}), at least for the majority of
objects, and hence a variety of techniques can be employed to identify
AGN from star-forming galaxies.

A common approach adopted to identify AGN at mid-IR wavelengths is to
use color-color diagnostics; see Table~2 of \citet{pado17agn} for a
comprehensive list of different infrared color-color diagnostics. The
basic principle behind this approach is the same as that adopted for
unobscured AGN searches using UV--optical color-color diagnostics;
i.e.,\ exploiting differences between the SEDs of AGN from other
astrophysical source populations. However, the key advantage that the
mid-IR waveband provides over the UV--optical waveband is that both
obscured and unobscured AGN are selected. The first color-color
diagnostics in the infrared waveband were developed over 30 years ago
using broad-band data over 12--100~$\mu$m from the {\em IRAS}
observatory \citep[e.g.,][]{degr85iragn, low1988}. More recently,
mid-IR color-color diagnostic diagrams have been developed based
around the sensitive {\em Spitzer} and {\em WISE} observatories at
3--24~$\mu$m \citep[e.g.,][]{ster05, alon06, mate12xmmwise,
  assef2013}. Although powerful, these color-color diagnostics do have
some limitations: (1) the source needs to have a strong AGN component
to be selected and hence systems with weak AGN components
(e.g.,\ intrinsically weak AGN or AGN hosted in strongly star-forming
galaxies) are not easily identified and (2) the mid-IR colors of
high-redshift ($z>$~2--3) star-forming galaxies can be similar to
those of AGN and hence ``contaminate'' the AGN selection parameter
space in the color-color diagrams. These limitations can be mitigated
to a large extent by further requiring a minimum flux threshold (which
removes the high-redshift star-forming galaxies, which are faint at
mid-IR wavelengths) or by combining mid-IR data with far-IR data to
construct broad-band infrared SEDs to search for weaker AGN components
in the mid-IR waveband (see Section \ref{sec:irradio}). In nearby systems, high spatial resolution mid-IR observations can distinguish a compact nucleus from more extended star formation and so potentially identify weaker AGN \citep[e.g.,][]{sieb08mir}.  Overall, the
completeness of the AGN selection in the mid-IR waveband can be high,
but the reliability of the AGN selection is modest and
depends on how the aforementioned caveats are handled.

A surprising result from the analysis of the mid-IR continuum of AGN
is that obscured AGN have broadly similar mid-IR SEDs to unobscured
AGN, both when using measurements on the scale of the whole galaxy (e.g., \citealt{buch06, mate12xmmwise, hick17qso}; see {\bf Figure
  \ref{fig:spectrum}}) and for spatially-resolved measurements on $\sim$pc scales \citep[e.g.][]{ramo11irsed, asmu14mir}. This is in disagreement with that expected from
the basic unified AGN model, since the presence of obscuration should
suppress the mid-IR emission from an obscured AGN (with greater
suppression of the shorter-wavelength emission). Indeed, it is often
not possible to reliably distinguish between an obscured and
unobscured AGN on the basis of just the mid-IR colors, and optical or
X-ray data are required to determine whether the AGN is obscured or
unobscured \citep[e.g.,][]{barm06, hick07abs}. These findings are a
key driver behind the idea that the obscuring dust in AGN is not
distributed in a smooth torus but is clumpy and allows for the mid-IR
emission to escape from the torus without being obscured
\citep[e.g.,][]{netz15unified, ramo17nucl}; see Section
\ref{sec:torus}. Furthermore, high spatial resolution mid-IR imaging has also revealed that a large fraction of the mid-IR emission from
some obscured AGN is produced from dust in the polar regions rather than the torus (e.g., \citealt{raba09ngc1068, asmu16polar}; see {\bf Figure~\ref{fig:structure}} and Section~\ref{sec:extended}).

\subsubsection{Spectroscopic techniques}

\label{sec:mirspec}

Strong discrimination between AGN and star-forming galaxies can be
achieved using mid-IR spectroscopy. The ``hotter'' radiation field
from the accretion disk, when compared to that from star-forming
regions, means that the detection of high excitation emission lines
(e.g.,\ [Ne~VI]~7.6~$\mu$m; [Ne~V]~14.3~$\mu$m; see
\citealt{spin92iragn}) provides an almost unambiguous identification
of AGN activity. However, the equivalent width of these
high-excitation emission lines is often low when compared to emission
lines at UV--optical wavelengths, making them sometimes challenging to
detect. Despite this, the low optical depth at mid-IR wavelengths
\citep[e.g.,][]{drai03dust} facilitates the detection of emission
lines in heavily dust-obscured regions where the UV--optical
signatures of AGN activity are extinguished \citep[e.g.,][]{saty08irs,
  goul09irs,pere10mirlines}. Overall, this technique provides an
obscured AGN identification approach with high completeness and
reliability; however, the lack of large-scale mid-IR spectroscopic
facilities mean that the sample sizes are currently small when
compared to those available from UV--optical spectroscopy.

The mid-IR spectra of AGN and star-forming galaxies also differ in the
strength of the broad-band spectral features due to polycyclic
aromatic hydrocarbon (PAH) molecules \citep[e.g.,][]{tielens2008}. PAHs
appear to be ubiquitous in the interstellar medium of galaxies
\citep[e.g.,][]{peet04pah, smit07pah} and are broadly correlated with
the star-formation component of galaxies. As a consequence, the
equivalent width of the PAH features can be used to assess the relative strength or
weakness of emission from the AGN in the mid-IR waveband. An advantage of
this approach is that lower resolution mid-IR spectroscopy can be
employed than that required for the detection of emission lines since
the PAH features have large equivalent widths
\citep[e.g.,][]{genzel1998,pope2008}. Weak AGN components can be
identified with this approach and it is particularly effective when
combined with mid-IR--far-IR photometry to provide a broader
wavelength baseline to constrain the strength of both the AGN and
star-formation components; see Section \ref{sec:irradio}.

Mid-IR spectroscopy can provide a reliable route to measure the amount
of obscuration towards the AGN. The primary spectroscopic diagnostic
is a strong absorption feature due to Si-based dust grains at 9.7 and
18~$\mu$m \citep[e.g.,][]{drai84dust}. The depth of the Si absorption
feature provides an estimate on the amount of obscuration towards the
mid-IR emitting region of the AGN. Obscured AGN are often found to
have Si absorption features while unobscured AGN typically have Si
emission features \citep[e.g.,][]{hao07irs,
  hatz15silicate,alon16irspec}. Overall a broad correlation is found
between the strength of the Si absorption and the absorbing column
density measured using X-ray data
\citep[e.g.,][]{shi06silicate,honi10torus}. The presence of strong Si
absorption is therefore sometimes taken as an indicator for a obscured
AGN, revealing potentially extremely obscured AGN that lack AGN
signatures at other wavelengths \citep[e.g.,][]{iman07irs,
  geor11silicate}. However, not all obscured AGN have strong Si
absorption, including perhaps half of the Compton-thick AGN
population, and the origin of the Si absorption feature often appears
to be due to dust in the host galaxy rather than the torus
\citep[e.g.,][]{goul12comp}.

\begin{textbox}[t]\section{COMMON FAR-IR--RADIO SELECTION CRITERIA FOR OBSCURED AGN}

Commonly used criteria for identifying AGN in this waveband include:

\begin{itemize}
\item Significant AGN contribution from fitting of AGN and galaxy templates to the mid-IR--far-IR SED; \item Large ratio of high-excitation to low-excitation CO lines or the detection of dense gas tracers (i.e. HCN, HCO+); \item High observed radio power (i.e. $P_{\rm 1.4 GHz}> 10^{25}$ W Hz$^{-1}$); \item Flat radio spectral index; and \item Excess of radio emission beyond what that predicted for star formation.
\end{itemize}

Due to low optical depth in the radio, most criteria to classify AGN as obscured rely on other wavebands after identification in the radio, but one technique is the detection of absorption from neutral hydrogen determined through the 21-cm line.
\end{textbox}

\subsection{Selection of obscured AGN at far-infrared--radio wavelengths}

\label{sec:irradio}

AGN produce emission across a broad range of wavelengths and therefore
obscured AGN can be selected at wavebands not explored so far in this
review; see {\bf Figure \ref{fig:sed}}. The far-IR--radio waveband, in
particular, provides the potential for many significant advances over
the selection of obscured AGN at other wavelengths, principally
because the optical depth is very low at these wavelengths
\citep{hild83submm}, allowing for even the most heavily obscured AGN
missed in the X-ray and mid-IR wavebands to be selected. However, the
full potential of these wavebands for the selection of obscured AGN is
yet to be realised due to the relatively modest sensitivities of
current facilities. A summary of far-IR--radio methods for identifying obscured AGN is given in the sidebar on ``Common far-IR--radio selection criteria for obscured AGN'', and we discuss them in detail below. 

\subsubsection{Far-infrared--millimeter wavelengths}

\label{sec:farir} The continuum emission at far-IR--millimeter wavelengths
(30~$\mu$m--10~mm) from the majority of AGN is dominated by dust
heated from star formation in the host galaxy. This limits the
effectiveness of AGN identification on the basis of far-IR--millimeter
photometry alone. However, when the mid-IR photometry is combined with
the far-IR--millimeter photometry to construct the broad-band
IR--millimeter SED, the signature of an AGN component can be
identified through fitting the SED with AGN and star-forming galaxy
templates or models. This approach can potentially identify weaker AGN
components than that achieved through mid-IR photometry alone
\citep[e.g.,][]{pope2008,sajina2012,delm16qso} and is particularly
effective when combined with mid-IR spectroscopy (see Section
\ref{sec:mirspec}). 

Obscured AGN can also be identified using dense molecular gas tracers
with (sub)-millimeter spectroscopy (e.g.,\ HCN and HCO+;
\citealt{gao04hcn, aalt15hcn, iman16alma2}). These emission lines from dense molecular gas
 are radiatively excited by mid-IR photons and can therefore
reveal the presence of an obscured AGN. A similar technique uses observations of CO lines, for which the relative strengths of the rotational transitions depend on the excitation mechanism, and a high ratio of high-excitation to low-excitation CO lines can indicate heating from an AGN \citep[e.g.,][]{rose15her, ming18ngc34}. Since the optical depth is
very low even for Compton-thick levels of absorption at (sub)-millimeter
wavelengths \citep{hild83submm}, heavily obscured AGN missed at other
wavelengths can be identified using (sub)-millimeter spectroscopy
\citep[e.g.,][]{aalt15hcn, iman16alma2}; however, dilution from star
formation within the host galaxy will weaken the AGN signature. This approach offers great potential to extend our census of obscured AGN, although given the sensitivity of current facilities, the majority of
obscured AGN searches with this technique are limited to comparatively
nearby systems.

\subsubsection{Radio wavelengths}

\label{sec:radio}

The identification of AGN in the radio waveband ($\approx$~0.01--30~m)
has a long history going back to the first detected quasars
\citep[e.g.,][]{baad54radio1, schm63}. The dominant physical process
for AGN in the radio waveband is synchrotron emission, which can be
due to processes related to the accretion disk and/or large-scale
radio jets (see \citealt{pado16radio,tadhunter16} for a recent
review). The optical depth for radio emission is very low and so radio selection can identify very heavily obscured sources \citep[e.g.,][]{wilk13radio}, although
synchrotron self absorption can occur in compact radio-emitting
sources and H~I absorption is seen at 21~cm (the spin-flip
transition). However, AGN are not the only extragalactic source
population that can produce significant radio emission: star-forming
galaxies can also be bright in the radio band \citep[e.g.,][]{cond92radio}.

At the highest radio luminosities (e.g.,\ $>10^{25}$~W~Hz$^{-1}$ at
1.4~GHz) AGN are uniquely distinguished from star-forming galaxies:
these sources are often referred to as ``radio-loud'' AGN and comprise
a minority of the overall AGN population, which is predominantly radio
quiet \citep[e.g.,][]{pado16radio,tadhunter16}. At lower radio
luminosities AGN cannot be reliably distinguished from star-forming
galaxies on the basis of luminosity alone. However, since the radio
luminosity from star formation is tightly correlated with the far-IR
luminosity \citep[e.g.,][]{helo85, cond92radio}, AGN can be identified
by selecting sources that produce excess radio emission over that
expected from star formation \citep[e.g.,][]{donl05radio,
  delm13excess}. When multi-frequency radio data are available, the
radio spectral slope can also be used to identify AGN activity: a flat
radio spectral slope ($\alpha<0.5$; e.g.,\ \citealt{pado16radio})
indicates a compact source (synchrotron self absorbed) and therefore
an AGN with a steep radio spectral slope can be due to either AGN
activity or star formation. Higher spatial resolution data
(e.g.,\ very long baseline interferometry data) can also be used to
identify the presence of AGN activity over star formation: an
unresolved radio core, radio jets, and radio lobes indicate the
presence of an AGN \citep[e.g.,][]{pado16radio,tadhunter16}. Radio
wavelengths can therefore provide reliable obscured AGN selection with
high completeness, particularly at high radio luminosities. However,
the reliability of the obscured AGN selection in the radio band
decreases towards lower luminosities and depends on the luminosity of
the radio core and the multi-wavelength data available.

The radio waveband can also provide reliable absorption
measurements. The identification of the neutral H~I absorption feature
at 21~cm provides a measurement of the H~I column density towards the
radio-emitting source. Current studies suggest a connection between
the neutral H~I absorbing column at 21~cm with the absorbing columns
measured in the X-ray band \citep[e.g.,][]{osto16hi, moss17hi};
however, the current sample sizes are small and greater progress will
be made with future radio facilities such as the Square Kilometer
Array (SKA); see Section 6.1.5.

\subsection{Multi-wavelength identification and a comparison of selection methods}

\label{sec:multiwave}

No single waveband provides a complete and reliable selection of AGN with current facilities. The low optical depth at
mid-IR--radio wavelengths ensures weak obscuration effects and hence
high obscured-AGN completeness, although contamination of the AGN
emission from the host galaxy reduces the reliability of the AGN
selection when the host galaxy is bright with respect to the AGN. The
reliability of the AGN selection at mid-IR--radio wavelengths can be
significantly improved with high spatial resolution observations,
where the relative contrast between the AGN and the galaxy will be
higher, and from spectroscopic observations, where the identification
of emission lines and solid-state features provide constraints on the
relative strength of the AGN and host-galaxy emission
processes. However, the availability of high spatial resolution and
spectroscopic observations at mid-IR--radio wavelengths is limited
when compared to broad-band photometric data. By comparison,
UV--optical spectroscopy is often more readily available (thanks to ground-based multi-object spectrographic instruments) and can
select AGN with good reliability and completeness but is biased
against identifying AGN that reside in galaxies that are either
strongly dust obscured or bright when compared to the AGN. The X-ray
waveband has low optical depth, particularly in the hard band, and the
host-galaxy contamination is low, allowing for the reliable selection
of obscured AGN except for low-luminosity systems; however,
Compton-thick AGN are weak and can be challenging to identify at X-ray
energies.

\begin{figure}[t]
\includegraphics[width=\textwidth]{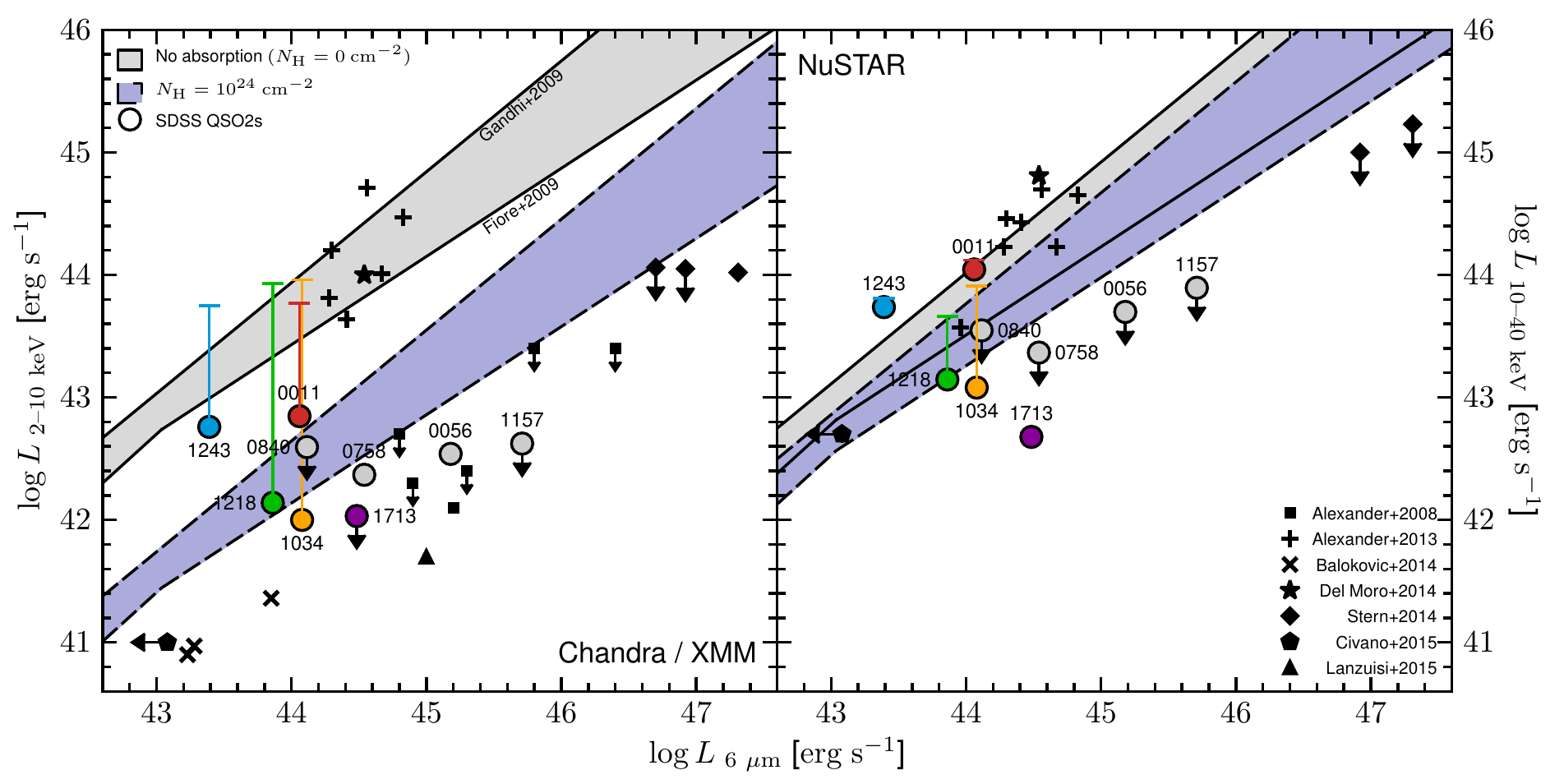}
\caption{X-ray luminosity versus 6~$\mu$m AGN luminosity for AGN
  explored in a variety of studies using (left) {\it Chandra} or {\it
    XMM-Newton} soft-band (2--10~keV) constraints and (right) {\it
    NuSTAR} hard-band (10--40~keV) constraints. The symbols indicate
  the observed X-ray luminosities and the solid vertical lines
  indicate the intrinsic (i.e.,\ corrected for absorption) X-ray
  luminosities for the sources with absorbing-column measurements. The
  grey shaded region indicates the range in intrinsic X-ray--6~$\mu$m
  AGN luminosity relationships between Fiore et~al. (2009) and Gandhi
  et~al. (2009) and the blue shaded region indicates the same
  relationships but where the X-ray luminosity is absorbed by a column
  density of $N_{\rm H}=10^{24}$~cm$^{-2}$. Taken from
  \citet{lans15nustarqso}, courtesy of G.~Lansbury.}
\label{fig:irxray}
\end{figure}

Due to the limitations in the identification of obscured AGN in any
given waveband, a more complete selection will be achieved
from a combination of multi-wavelength identification approaches. For
example, an AGN selection approach that combines X-ray, infrared, and
radio data (e.g.,\ such as that available for many of the blank-field
extragalactic survey areas) will reduce the identification biases of
any individual waveband. Such an approach would allow for the
identification of X-ray detected Compton-thin AGN even in strongly
star forming galaxies, the identification of potential infrared-bright
Compton-thick AGN from the weak or non detection of X-ray emission,
and the identification of radio-bright AGN that are heavily obscured
in both the X-ray and infrared wavebands \citep[e.g.,][]{hick09corr,
  delm13excess, delm16qso}. In {\bf Figure \ref{fig:irxray}} we
demonstrate the complementarity of infrared and X-ray data in
identifying potential Compton-thick AGN from the detection of bright
mid-IR emission from the AGN with weak or undetected X-ray emission
\citep[e.g.,][]{alex08compthick, vign10qso2, lans15nustarqso,
  lans17nustar}. Combining these multi-wavelength data with spectroscopic
observations would provide an even more complete selection of obscured
AGN.

However, despite the obvious advantage of a multi-wavelength approach
in terms of providing a more complete selection of AGN, the relative
simplicity of the single waveband approach does have a key attribute:
a simple selection function. The selection function is the
quantification of the sensitivity and identification biases and will
be much simpler for a single waveband approach than for a
multi-wavelength approach. Therefore, if the effect of obscuration and
host-galaxy dilution is well understood in the selected waveband then
it can be used to {\em model} the data and {\em infer} the properties
of the overall AGN population (i.e.,\ taking into account the AGN not
selected due to sensitivity and identification biases). We have
emphasized {\em model} and {\em infer} here since this approach is
model based (e.g.,\ it makes assumptions about the AGN and host galaxy
properties, an example being the distribution of AGN absorbing column
densities) and the complete AGN population is inferred rather than
directly identified. For complex selection functions that depend on
several variables, an effective technique is to simulate the overall
AGN population and then apply the same identification procedure to the
simulated data; this technique is commonly referred to as ``forward
modeling''; see Section \ref{sec:models}. The selection function is
then the difference between the input and the output (i.e., the
overall AGN population and the subset of the AGN population that are
identified).

Multi-wavelength and single-waveband identification approaches are
therefore complementary. The multi-wavelength approach can construct a
(near) complete census of AGN in, for example, a given volume down to
a given luminosity. It requires more extensive data than the
single-waveband approach and will likely be limited in the
volume--luminosity parameter space that it can cover. However, the
knowledge of the AGN properties gained from this approach can then
guide the modeling required to infer the overall AGN population from a
single waveband approach, which requires less extensive
multi-wavelength data and can extend over larger regions of parameter
space. One application of this approach is constraining the properties of the
Compton-thick AGN population. X-ray data are essential in identifying
Compton-thick AGN as it is required to constrain the absorbing column
density. However, a flux-limited X-ray survey is biased against
detecting Compton-thick AGN due to the suppression of the X-ray
emission and, therefore, significant corrections are required to infer
the overall Compton-thick AGN population (e.g.,\ an observed fraction
of $\approx8^{+1}_{-2}$\% vs an intrinsic fraction of
$\approx27\pm4$\% from the hard-band selected {\it Swift}-BAT survey;
Ricci et~al. 2015; see {\bf Figure \ref{fig:nhdist}}). The
multi-wavelength identification and characterisation of all AGN,
including measuring (or inferring) the absorbing column densities,
within a given volume \citep[e.g.,][]{goul09irs, annu15, annu17cthick}
is therefore essential to validate the modeling assumptions adopted in
the single waveband approach. Furthermore, since all of the AGN are
identified and characterised in the volume-limited study, it will
include systems that have abnormal properties or are intrinsically
weak in any given waveband, extending our understanding of the overall
AGN population.

\section{THE DEMOGRAPHICS OF THE OBSCURED AGN POPULATION}

\label{sec:demographics}

\begin{figure}[t]
\includegraphics[width=3.5in]{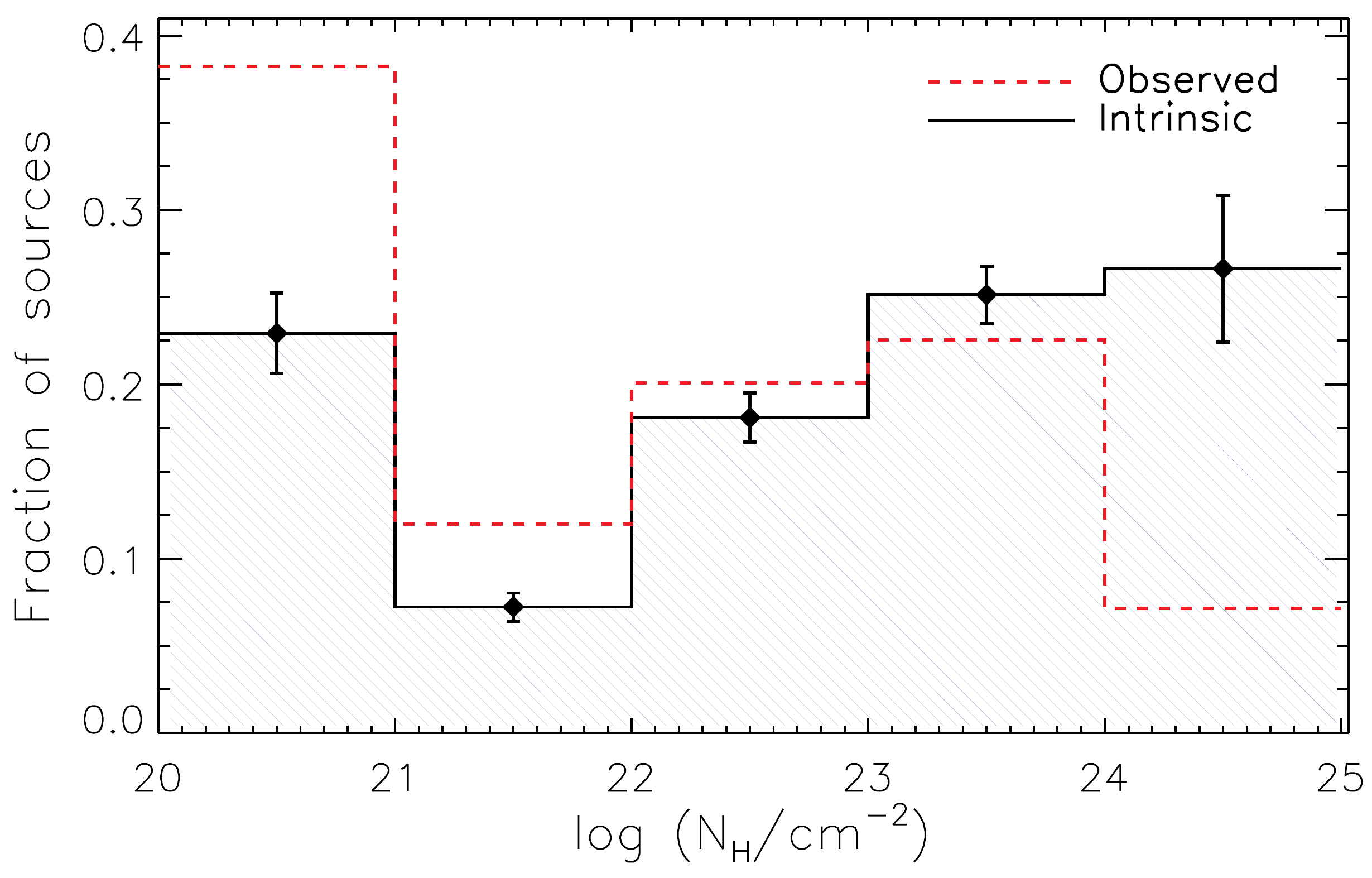}
\caption{Distribution of absorbing column densities ($N_{\rm H}$) for local AGN from the {\em Swift}-BAT survey. Both
  the observed distribution (red dashed line) and the intrinsic
  distribution after correction for selection effects (black solid line) are shown. A comparison between the
  two distributions demonstrate the large corrections required to
  determine the intrinsic fraction of Compton-thick AGN with $N_{\rm
    H}$~=~$10^{24}$--$10^{25}$~cm$^{-2}$. Figure from \citet{ricc17batagn}, courtesy of C.~Ricci.}
\label{fig:nhdist}
\end{figure}

As a prelude to the next section on the physical nature of obscuration in
AGN we provide a brief overview of the observed demographics of
obscured AGN activity. In this section we discuss three key aspects:
(1) the distribution of AGN absorbing column densities, including the
fraction of AGN that are Compton thick, (2) the luminosity dependence
of obscuration, and (3) the redshift dependence of obscuration. We
will mostly focus our discussion on results from X-ray observations
since they provide an efficient AGN selection and yield one of the
most reliable absorption measurements; however, we will note
similarities and discrepancies with results obtained at other
wavelengths.

The majority of the AGN population are obscured: they dominate both
the number density and luminosity density of accretion onto SMBHs
\citep[e.g.,][]{ueda14cxb, aird15xlf, buch15xlf}. A common way to
characterise the amount of obscuration is to construct the
distribution of absorbing column densities (often called the $N_{\rm
  H}$ distribution). In {\bf Figure \ref{fig:nhdist}} we show an
example $N_{\rm H}$ distribution for AGN in the local Universe
detected at 14--195~keV from the {\it Swift}-BAT all-sky survey
\citep{ricc17batagn}. The absorbing column densities have been
measured using the {\it Swift}-BAT data in combination with X-ray data
in the soft band and both the observed and {\it intrinsic} (i.e.,\ the
inferred $N_{\rm H}$ distribution after correcting for the selection
function; see Section \ref{sec:multiwave}) distributions are
plotted. The intrinsic fraction of obscured AGN ($N_{\rm
  H}>10^{22}$~cm$^{-2}$) from this study is $70\pm5$\% and the
intrinsic fraction of Compton-thick AGN ($N_{\rm
  H}=10^{24}$--$10^{25}$~cm$^{-2}$) is $27\pm4$\% \citep[e.g., also
  see][]{akylas2016,koss2016}. Note that the absence of heavily
Compton-thick AGN with $N_{\rm H}>10^{25}$~cm$^{-2}$ is not intrinsic
but is due to the low sensitivity of X-ray observations to such large
amounts of obscuration and therefore the Compton-thick fraction is
actually a lower limit. An effective technique that can be adopted to
identify heavily Compton-thick AGN is to combine mid-IR and X-ray
observations to select AGN that are bright in the mid-IR waveband but
weak or undetected at X-ray energies; see Section \ref{sec:multiwave}
and {\bf Figure \ref{fig:irxray}}.

\begin{figure}[t]
\hspace{-0.2in}
\begin{minipage}{2.6in}
\includegraphics[width=2.6in]{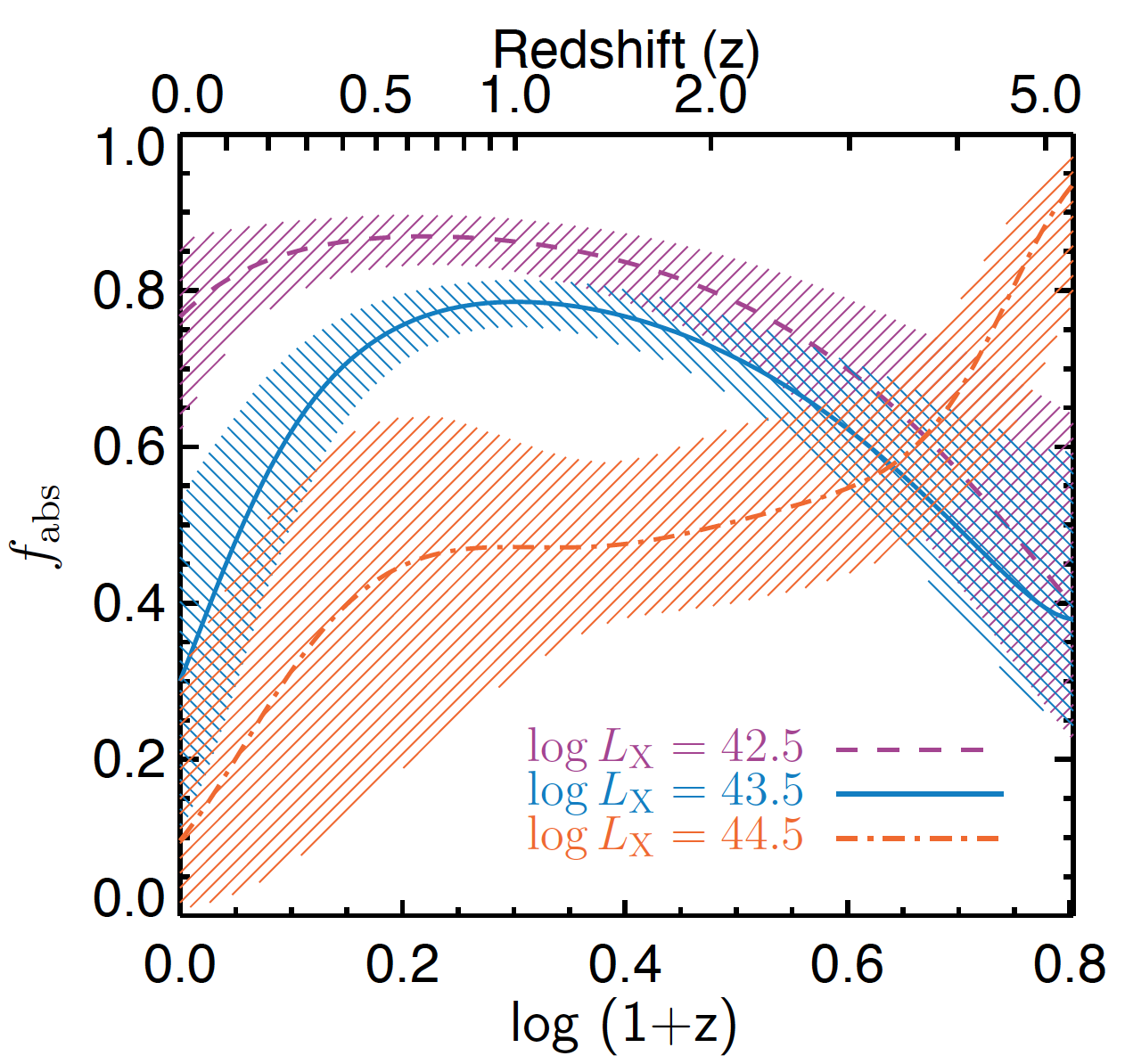}

\end{minipage}
\begin{minipage}{2.6in}
\includegraphics[width=2.6in]{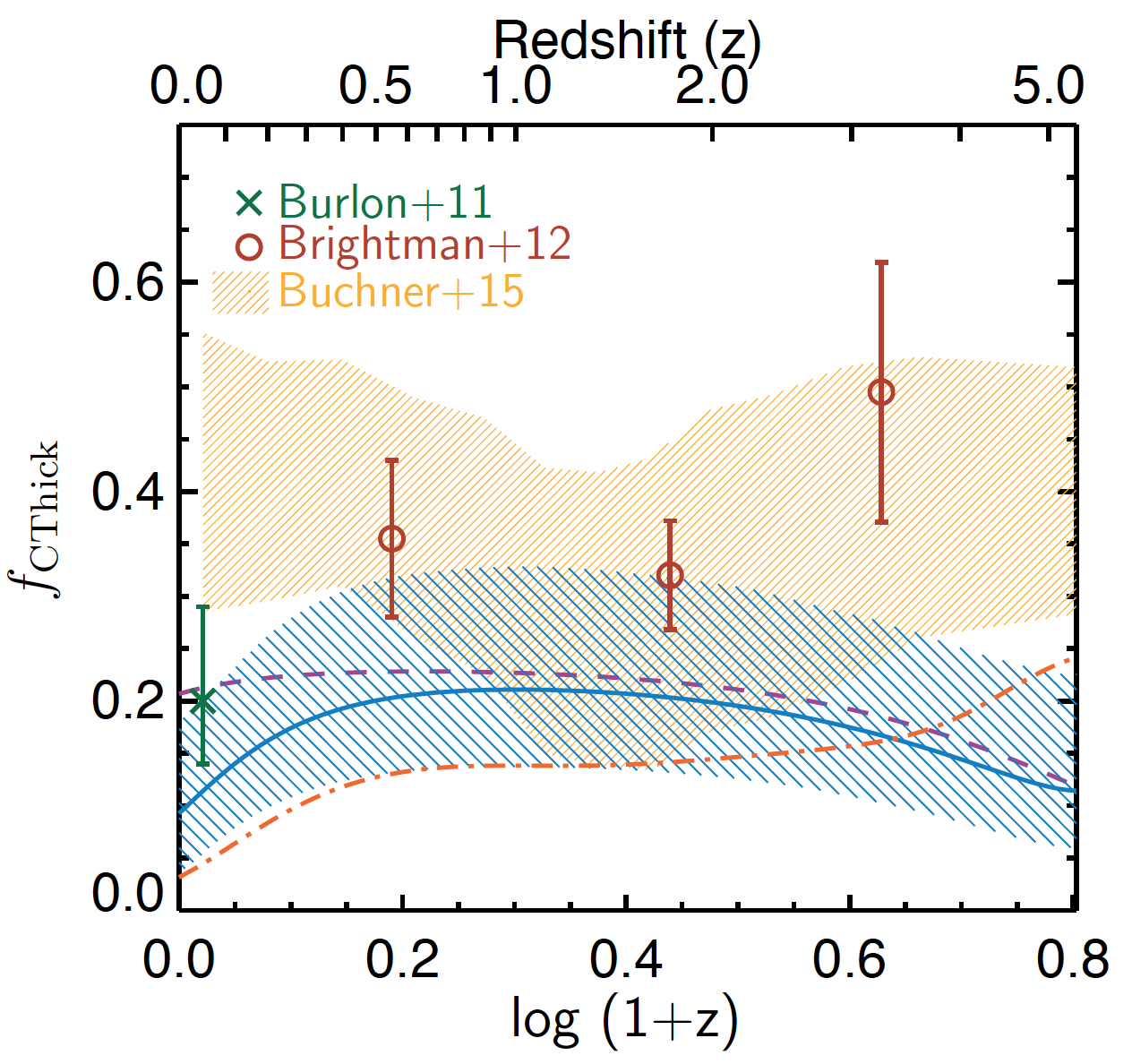}
\end{minipage}
\caption{(Left) Fraction of AGN that are X-ray absorbed but Compton
  thin vs redshift for three different rest-frame 2--10~keV
  luminosities (colored coded, as indicated) for the best-fitting
  model presented in Aird et~al. (2015b). (Right) Fraction of AGN that
  are Compton-thick vs redshift at a rest-frame 2--10~keV luminosity
  of $10^{43.5}$~erg~s$^{-1}$ (blue curve and shaded region), compared
  to other studies (as indicated) for the best-fitting model presented
  in Aird et~al. (2015b). The magenta and orange curves show the
  Compton-thick AGN fraction vs redshift at a rest-frame 2--10~keV
  luminosity of $10^{42.5}$~erg~s$^{-1}$ and $10^{44.5}$~erg~s$^{-1}$,
  respectively. Figures from \citet{aird15xlf}, courtesy of
  J.~Aird.}
\label{fig:aird15}
\end{figure}

The fraction of AGN that are obscured appears to be a function of both
AGN luminosity and redshift. The first tentative evidence of this
result was found over 25 years ago \citep{lawr91} and has since been
confirmed by many studies in the X-ray, optical, infrared, and radio
wavebands \citep[e.g.,][]{simp05agn, maio07cover, trei08obsqso,
  luss13obscagn, mate17torus}. An example of the obscuration
dependence with AGN luminosity is shown in {\bf Figure
  \ref{fig:fobsc}} for X-ray detected AGN over the broad redshift
range of $z=$~0.3--3.5 \citep{merl14agnobs}. An interesting attribute
of this chosen example is that the distinction between obscured and
unobscured AGN is made using both X-ray data and optical
spectroscopy. However, regardless of whether the AGN are classified
as obscured in the X-ray or optical waveband, there is a clear
decrease in the obscured AGN fraction towards higher luminosities. The
decrease implies an increase in the opening angle of the torus with
luminosity and therefore a decrease in the dust covering fraction, a
result that is often interpreted as due to a ``receding torus'' (see
Section \ref{sec:torus}). We note that there is considerable variation
between studies in the strength of the decrease in the obscured AGN
fraction with luminosity (e.g.,\ see Figure~28 of \citealt{toba14cover}
for a compilation of multi-wavelength results). Several factors are
likely to contribute to the variation in results, including (1) the
method adopted to distinguish between obscured and unobscured AGN, (2)
the waveband used to select the AGN (i.e.,\ the optical depth and
sensitivity towards obscuration), and (3) the range of parameter space
used to select the AGN (e.g.,\ across the redshift--luminosity--mass
plane). Furthermore, as we discuss in Section \ref{sec:torus}, the
primary driver of the obscured AGN fraction may be Eddington ratio
rather than luminosity.

In general the evidence for evolution in the obscured AGN fraction
with redshift is less secure than with luminosity. An example study is
shown in {\bf Figure \ref{fig:aird15}}, which presents the evolution
in the X-ray absorbed fraction of Compton-thin and Compton-thick AGN
with redshift \citep{aird15xlf}. These constraints were derived from
fitting a model to the measurements of the evolving X-ray luminosity
functions of X-ray absorbed, X-ray unabsorbed, and Compton-thick AGN
using data from deep X-ray surveys; the X-ray luminosity function is
the measurement of the space density of AGN as a function of X-ray
luminosity, taking into account the X-ray selection function. There is
considerable uncertainty in the measured redshift evolution in the
obscured AGN fraction. However, overall, the studies broadly agree
that the obscured fraction of distant AGN is at least comparable with
that found locally and may increase with redshift \citep[e.g.,][]{ueda14cxb, buch15xlf}, which could be
driven by the increase in
the star-formation rate and cold-gas fraction of galaxies with
redshift (see \citealt{mada14sfz} for a recent review).

\section{THE PHYSICAL NATURE OF OBSCURATION IN AGN}

\label{sec:physics}

\begin{figure}[t]

\includegraphics[width=0.9\textwidth]{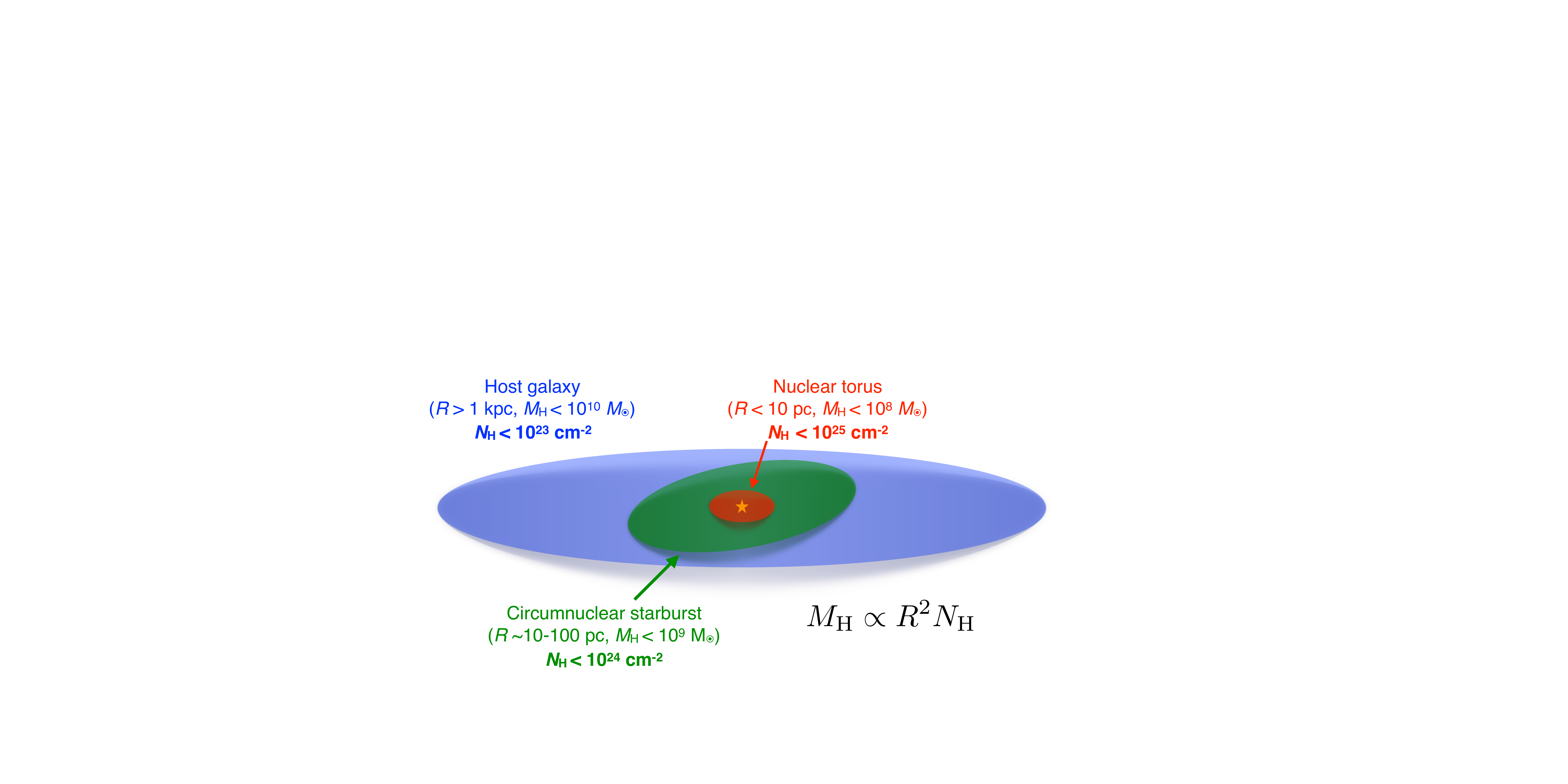}

\caption{A schematic representation of the different scales of AGN obscuration considered in Section~\ref{sec:physics}, for a Milky Way-type galaxy in the local Universe. For the vastly simplified assumption of constant density, for a given gas mass $M_{\rm H}$, the typical column density $N_{\rm H}$ toward the nucleus decreases with size scale as $R^{-2}$; the heaviest obscuration thus tends to occur on the smaller scales, although larger-scale Compton-thick obscuration can occur in discrete events such as galaxy mergers (Section~\ref{sec:evolution}) or at high redshifts where the gas fraction in galaxies is large (Section~\ref{sec:highz}).}
\label{fig:scales}
\end{figure}

The previous sections illustrate the ubiquity of obscuration in AGN
and the diversity of the associated observational signatures. As we discuss below, AGN obscuration is intimately connected to  both the fueling of the SMBH (through inflows of gas) and ``feedback'' (produced by the radiative and mechanical power of the AGN). To understand this connection, we require knowledge of the nature of the obscuring material:
the scales, densities, composition, kinematics of the obscuring
clouds, and the physical processes that produce them. A number of
studies have treated AGN obscuration as dominated by a single regime
(most often on the scale of a torus; e.g., \citealt{davi15torus,
  mate16torus, mate17torus}), but it is increasingly clear that
obscuration can occur on a range of scales and physical
conditions. Furthermore, time-varying obscuration has been invoked in
models of SMBH-galaxy co-evolution to explain the SMBH-galaxy
relationships and the observed connection between AGN and starburst
activity \citep[e.g.,][]{dima05qso,hopk08frame1,alex12bh}. In this section
we will focus on three main regimes of obscuration illustrated in {\bf Figure~\ref{fig:scales}}: (1) the nuclear
``torus'' posited by AGN unification models, (2) circumuclear gas
associated with central starbursts, and (3) galaxy-scale material
associated with galaxy disks and mergers.

An important consideration when comparing these different regimes is the
characteristic obscuring column densities that can be associated with
each scale. In the simplified case of a hydrogen cloud with constant
number density $n_{\rm H}$ distributed over a sphere of radius $R$, the
column density will be $N_{\rm H} = n_{\rm H} R$ and the gas mass will
be, in terms of the mass of the hydrogen atom $m_{\rm H}$, $M_{\rm gas} = m_{\rm H} n_H \frac{4}{3} \pi R^3$, such that $M_{\rm H}
\propto N_{\rm H} R^2$. Thus for a given mass of gas, $N_{\rm H} \propto
R^{-2}$, suggesting that the highest column densities will occur on
relatively small scales and therefore that only modest amounts of
obscuration can be expected over larger scales for a reasonable mass of
gas, as illustrated in the schematic in {\bf Figure~\ref{fig:scales}} (see also \citealt{buch17galobsc}).
(We note however that instabilities and disturbances such as galaxy
mergers can significantly increase the gas density on $\sim$100 pc to kpc scales, and temporarily
produce larger columns along some lines of sight, as discussed in Section~\ref{sec:evolution}.) In
what follows, we will discuss the different regimes of obscuration in
the context of observational and theoretical constraints on the typical
values of $N_{\rm H}$.

\subsection{The nuclear torus and the unified AGN model}

\label{sec:torus}

As discussed in Section \ref{sec:intro}, in the physical model for
the AGN central engine comprises of a small-scale, broadly
axisymmetric structure of dust and gas that surrounds the SMBH,
accretion disk, and BLR clouds, and obscures them along some lines of
sight \citep[e.g.,][]{anto93, urry95, netz15unified}. This ``unified AGN model''
has been remarkably successful at explaining a number of properties of
individual AGN (such as the existence of hidden BLRs in some
Seyfert 2s; see Section \ref{sec:optspec}) and of the demographics of the AGN population as a whole
(such as the correlation between the X-ray luminosity produced by the
accreting SMBH and the mid-IR luminosity that is reprocessed in the
torus). However, in recent years it has become clear that the simplest
models of the torus that posit a smooth, symmetric ``donut''-like
structure are inconsistent with observations, as discussed in Section 2.3. Recent reviews by \citet{netz15unified} and \citet{ramo17nucl} give a comprehensive treatment
of the status of the unified AGN model and its successes and challenges;
here we will give a brief overview of some of the key points.

\subsubsection{The basic torus properties are well-constrained}
\label{sec:torusprop} It is well-established that the inner regions of
the obscuring torus are relatively compact ($<1$ pc), from near-
and mid-IR measurements of reverberation time lags
\citep[e.g.,][]{suga06reverb, vazq15torus} and spatially resolved
emission from dust using mid-IR interferometry \citep[e.g.,][]{lope16polar}. The radii of these mid-IR detected structures (assumed to be the AGN torus) closely follow a relationship with AGN luminosity of $r_{\rm torus}
\propto L^{1/2}$ that is remarkably consistent with the predicted
sublimation radius for graphite dust \citep[e.g.,][]{barv87dust,
burt13seyfir}) that is expected to represent the inner edge of the torus.  A natural scale for the outer edge of the torus is the gravitational sphere of influence of the SMBH
(within which the SMBH dominates the gravitational field), 
which will broadly correspond to a radius of $\sim$10 pc for nearby systems (e.g., see Section 2.2 of \citealt{alex12bh}). Observations of the outer regions of the torus have come from mid-IR imaging \citep[e.g.][]{asmu16polar} and studies using molecular lines \citep[e.g.,][]{garc16ngc1068}, although the precise outer edge
of the torus may be difficult to distinguish from a nuclear starburst
disk, as discussed in Section~\ref{sec:starburst}. The wide observed
range of obscuring $N_{\rm H}$ in X-ray studies of AGN suggest a range of
column densities through the torus, although the ubiquity of Fe K$\alpha$ reflection
features in AGN X-ray spectra (Section \ref{sec:xray}) suggests that in
general, AGN tori are Compton-thick along some lines of sight.

\subsubsection{The torus is clumpy} As discussed in Section ~\ref{sec:midir}, a broad range of evidence points to the
torus being highly inhomogeneous in density, temperature, and composition, so that its overall structure is clumpy rather than smooth. 
One important piece of observational evidence pointing toward a clumpy structure comes from high-resolution mid-IR imaging. Models for smooth tori consistently predict weaker mid-IR emission for edge-on (Type 2) systems due to the torus itself having a large
optical depth in the mid-IR, so that its emission is anisotropic \citep[e.g.,][]{frit06torus}. For
a clumpy torus, we expect to observe only the surfaces of the
optically-thick clumps, which can be illuminated deep within the torus due to the optically thin lines of sight through the gaps between the clumps. This scenario produces mid-IR emission that is much less
dependent on orientation \citep[e.g.,][]{nenk08torus, stal12torus}, in agreement with the remarkably tight observed relationship between X-ray and mid-IR
luminosities that is consistent for both Seyfert 1 and 2 galaxies
\citep[e.g.,][]{gand09seyfir, garc16mir}.  Further evidence for a clumpy torus comes from observations of Si absorption features in the mid-IR spectrum (Section \ref{sec:mirspec}).
A common prediction of models for smooth tori viewed edge-on are deep absorption Si absorption lines \citep[e.g.,][]{frit06torus}. However a study of local
Compton-thick AGN by \citet[Section \ref{sec:midir}]{goul12comp} showed that deep Si absorption is most often associated with larger-scale structures (dust lanes or galaxy merger features) rather than a smooth, small-scale torus. By contrast, high-angular resolution nuclear spectra of face-on, isolated Type 2 Seyferts show shallower Si absorption features that can be naturally produced by clumpy torus models \citep[e.g.,][]{roch06circ, alon16irspec}.

\subsubsection{The torus can have a range of covering factors, with dependence on AGN properties} 

One key parameter of the torus is the opening angle, or equivalently,
covering factor ($f_C$). An estimate of $f_C$ can be obtained for
individual sources from detailed modeling of the X-ray spectrum
\citep[e.g.,][]{brig11torus} and studies of the ratio of reprocessed
(IR) to direct (optical or X-ray) AGN emission
\citep[e.g.,][]{toba14cover}, while the the average $f_C$ for an AGN
population can be inferred from the fraction of sources that are
obscured for given AGN parameters \citep[e.g.,][]{lawr91}.  Individual
AGN are observed with opening angles over the full range from 0 to 90
degrees \citep[e.g.,][]{mate16torus}, and sources with larger $f_C$ are
statistically more likely to be observed as obscured than unobscured
\citep[e.g.,][]{elit12unified}. Even for AGN of similar mass and
luminosity, a broad range of torus properties are observed
\citep[e.g.,][]{ramo11irsed, burt13seyfir}. 

Despite this broad diversity in the tori of individual AGN, there are general trends in
average $f_C$ with various AGN parameters.  It has long been observed
that the obscured fraction (and thus the average $f_C$) decreases with AGN luminosity (Section
\ref{sec:demographics}), which has been interpreted in terms of
receding torus models in which increasingly luminous AGN
progressively blow away more of the obscuring material \citep[e.g.,][]{lawr91}. Studies of optical and soft X-ray samples have suggested that
the obscured fraction drops as low as $\sim$10\% at the highest
luminosities \citep[e.g.,][]{lawr91, hasi08agn}. However, recent studies
including more sophisticated modeling of incompleteness and anisotropy
in the IR emission indicate a much weaker luminosity dependence, with
the obscured fraction remaining as high as 50\% even for the highest
luminosities \citep[e.g.,][]{stal16torus,mate17torus}. This suggests that while the inner radius of the torus increases with luminosity along with the dust sublimation radius (Section \ref{sec:torusprop}), the covering factor of the torus remains broadly constant at the highest luminosities.

It has recently been suggested that the key parameter determining $f_C$ may not be luminosity but  Eddington ratio ($L/L_{\rm Edd}$; \citealt{buch17galobsc, ricc17edd}), with $f_C$ limited by 
radiation pressure from the AGN acting on dust \citep[e.g.,][]{fabi08pres}. In a study of local hard X-ray selected AGN, \citet{ricc17edd} found that $f_C\approx0.8$ at $L/L_{\rm Edd}<0.02$ and then drops dramatically at higher $L/L_{\rm Edd}$, independent of AGN luminosity.  In this picture, the minimum $f_C$ of $\approx$30\% at $L/L_{\rm Edd}>0.5$ is set primarily by the covering factor of Compton-thick material along the equatorial plane of the torus.  A strong dependence of $f_C$ on $L/L_{\rm Edd}$ indicates that most of the obscuring material is within the gravitational sphere of influence of the SMBH, suggesting that (at least for the local AGN in their sample) that a compact torus-like structure is the dominant source of obscuration.

	\begin{figure}[t]

\includegraphics[width=\textwidth]{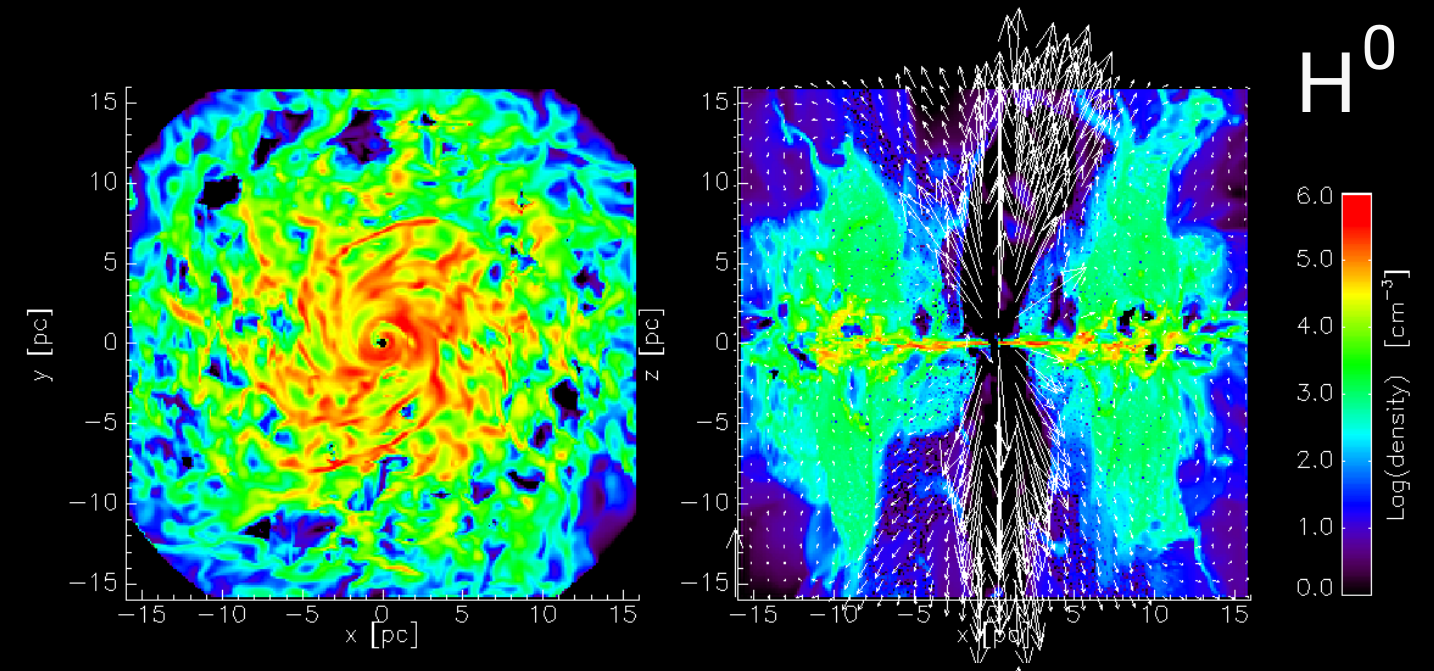}

\caption{Visualization of the density distribution in a hydrodynamic simulation of flows around a SMBH \citep{wada16torus}, shown face-on (left) and edge-on (right). The simulations illustrate a dynamical radiatively-driven ``fountain'' that can have obscuration and dust emission properties similar to those observed for some AGN. Figure from \citet{wada16torus}, courtesy of K.~Wada.}
\label{fig:wada}
\end{figure}

\subsubsection{The torus is dynamic} Any structure surrounding an accreting SMBH
exists in a complex environment of inflow and outflow. The dynamic nature of the torus is not captured by ad hoc models of smooth or clumpy tori (which are generally static with time) but appear naturally in hydrodynamical models of
gas flows around the SMBH. These can produce a variety of broadly axisymmetric
structures that may be associated with a torus, including a warped
accretion disk \citep[e.g.,][]{jud17disc} or the interaction of
inflowing gas with AGN- or starburst-driven winds to produce nuclear
structures with large scale heights (e.g., \citealt{wada16torus,
  hopk16agnfeed, honi17torus}; see {\bf Figure~\ref{fig:wada}}). Some observations of broad absorption line features in
AGN spectra have been interpreted as being viewed through axisymmetric
outflowing winds that could be interpreted as a torus-like structure
\citep[e.g.,][]{gall16torus}.

\subsubsection{The torus may extend in the polar direction, and to large
scales} 
\label{sec:extended} An axiomatic feature of most torus models is that the obscuring
gas and dust is broadly symmetric along the rotation axis of the
accretion flow. This is motivated the presence of ionization cones
observed in NLR gas \citep[e.g.,][]{zaka06host,fisc13nlr} and hidden BLRs in some Seyfert 2
galaxies (see Section~\ref{sec:optspec}).  However, recent
high-resolution observations of a handful of nearby AGN tori using
mid-IR interferometry revealed presence of substantial dust emission
along the polar direction on pc scales \citep[e.g.,][]{
honi12ngc424, tris14circ, lope16polar}. These polar structures can extend to larger scales as shown by imaging observations \citep[e.g.,][]{asmu16polar}. The
physical origin of these features is still unclear, but may be
associated with an AGN-driven outflow \citep[e.g.,][]{scha14torus}. In addition, 
recent observations have cast some doubt on the notion of a compact torus as being the
sole origin of reflected X-ray emission as is often assumed in modeling of
obscured AGN \citep[e.g.,][]{murp09mytorus, brig11torus}. Spatially
resolved {\em Chandra} observations have found evidence for Fe K$\alpha$
lines produced up to $\sim$kpc away from the nucleus
\citep[e.g.,][]{baue15ngc1068, fabb17kalpha}. Taken together, these
results suggest that emission features that have previously been
attributed to a compact, axisymmetric torus may often originate from gas and
dust with very different geometries. Obscuring material on larger scales
may be associated with nuclear starbursts, which are discussed in the
next subsection.

\subsection{Obscuration by nuclear starbursts}

\label{sec:starburst}

A starburst disk on $<$100 pc scales is a natural consequence of
the significant inflow of gas into the central regions of the galaxy
that is required to produce rapid accretion onto the SMBH
\citep[e.g.,][]{thom05rad, davi09ifu}. On the scales of the entire
galaxy, far-IR observations of AGN have shown that there is a relatively
weak correlation between the AGN luminosity and current (or recent) star
formation  \citep[e.g.,][]{rosa12agnsf, stan15agnsf}. However, these
relationships are found to become tighter when measured over smaller
spatial scales \citep[e.g.][]{dima12agnsf, esqu14agnsf}, confirming that
accreting SMBHs often have a substantial reservoir of gas within the
central 100 pc that can fuel a starburst disk. 

Such gas is generally kinematically decoupled from the larger galaxy
disk, and radiation pressure can cause the starburst disk to expand to a
large scale height \citep[e.g.,][]{thom05rad, hopk16agnfeed}. Sampling
all lines of sight, starburst disks can produce $N_{\rm H}$
distributions that are broadly consistent with observations of the AGN
population \citep{ball08agnsb, hopk16agnfeed, gohi17agnsb}. As per the discussion in Section~\ref{sec:torusprop}, Compton-thick obscuration in these
models is generally limited to small-scale structures ($<$1 pc for a
$3\times10^7$ $M_\odot$ SMBH) that are difficult to distinguish from a
torus. However, Compton-thin obscuration by starburst disks on
larger ($>$10 pc) scales may contribute significantly to the total
population of obscured AGN.

\subsection{Obscuration by galaxy-scale material}

\label{sec:galaxy}

In addition to structures directly related to accretion flows onto the
SMBH, obscuration can be produced by gas on the scales of the entire
galaxy ($>$kpc). In a cosmological context,
large-scale obscuration is common to models in which SMBH-galaxy co-evolution
is driven by galaxy mergers, whereby the gas flows onto the SMBH
are connected to galaxy-scale disturbances associated with
merger-driven torques \citep[e.g.,][]{dima05qso,
  hopk08frame1,alex12bh}. In this ``evolutionary'' picture, the
earliest phases of rapid SMBH growth are surrounded by powerful
starbursts and shrouded in dust clouds produced by the merger,
followed by a ``blowout'' due to radiative feedback of the AGN that
produces an unobscured quasar. Motivated by these theoretical
expectations, a number of observational studies have explored the
question of whether AGN obscuration can be associated with
galaxy-scale structures rather than a nuclear torus or starburst disk.

\begin{figure}[t]
\includegraphics[width=\textwidth]{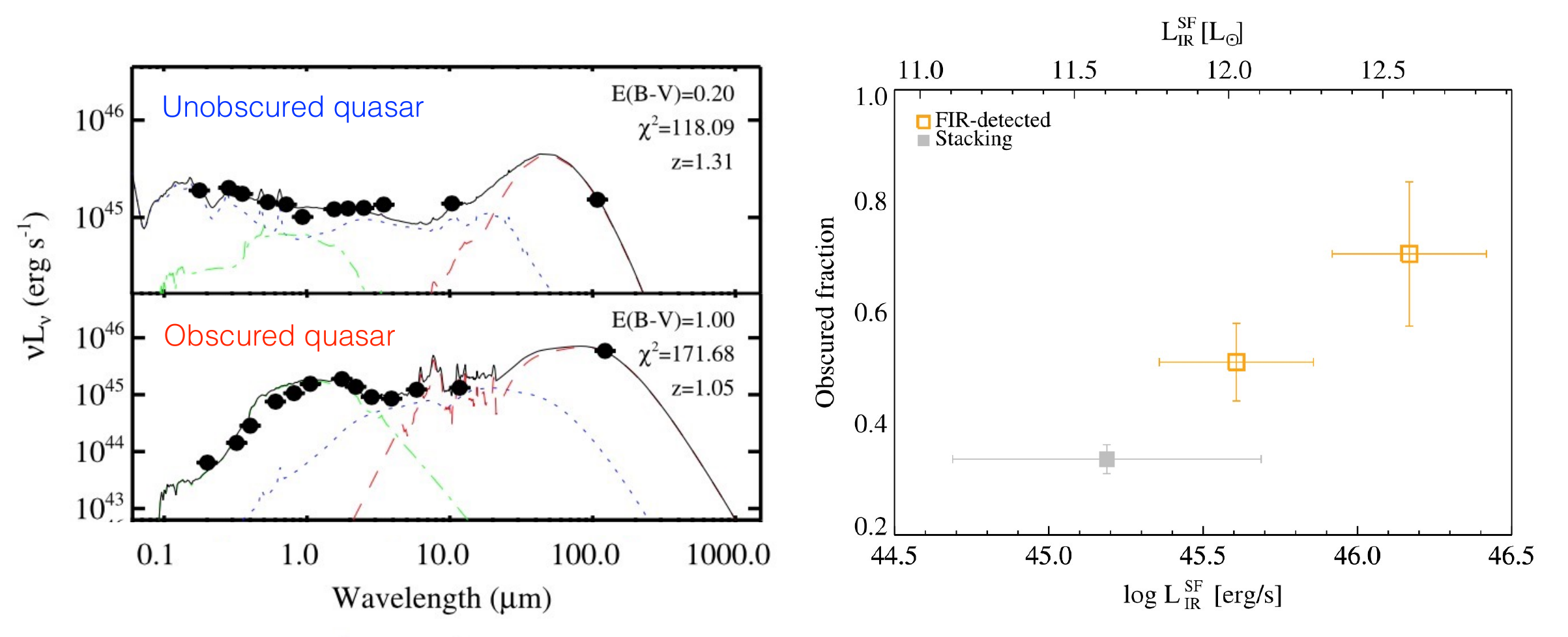}
\caption{The connection between obscuration and star formation in mid-IR luminous quasars, adapted from \citet{chen15qsosf}. The left panel shows fits to the optical--FIR SEDs (including {\em Herschel} data) of an unobscured and an obscured quasar identified using mid-IR and optical photometry \citep{hick07abs, chen15qsosf}. Obscured quasars in this sample exhibited stronger far-IR (cold dust) emission compared to their unobscured counterparts; the right panel shows that the fraction of AGN that are obscured rises significantly with far-IR luminosity, suggesting a connection between AGN obscuration and larger-scale star-forming dust, as discussed in Section~\ref{sec:galaxy}. Figures from \citet{chen15qsosf}, courtesy of C.-T.~Chen.}
\label{fig:chen}
\end{figure}

One approach to identifying galaxy-scale obscuration in AGN is to search
for links between obscuration and disturbed or merger morphologies of
the host galaxies. From an observational perspective, the merger-AGN
connection has been controversial, with some studies suggesting a strong
connection, others showing no relationship, and some suggesting a
dependence on AGN luminosity \citep[e.g.,][]{koss10batagn, trei12merge,
vill17merge}. Recent results indicate clearly that merging galaxies are
more likely to host AGN than isolated galaxies with otherwise similar
properties \citep[e.g.,][]{elli13merge, west17merge, goul17merge}, but
whether these mergers are associated with obscuration remains unsettled.
For low-luminosity nearby systems, the excess of AGN in mergers is
significantly stronger for AGN selected in the IR with {\em WISE} than for (presumably
less-obscured) optically-selected AGN \citep{saty14wisepairs}, although
the {\em WISE} color-selected AGN may suffer contamination from
low-metallicity starbursts \citep{hain16wise}. Studies of IR-selected
quasar hosts at $z\sim$~1--2 show no clear connection between merger
morphology and obscuration (\citealt{farr17obsc}), and similar results
were found for a X-ray selected AGN at $z\sim 2$ \citep{scha11xmorph,
koce12xmorph}.  However, recent studies report a possible
connection between galaxy mergers and Compton-thick AGN obscuration \citep{koce15xmerge, ricc17ctmerge}. While the obscuring material in these studies is usually modeled to have a small-scale torus geometry, in principle the characteristic X-ray features might be produced by reflection off clouds on larger scales associated with the merger (e.g., \citealt{leve02fek}, see Section~\ref{sec:extended}). For the population of reddened quasars (which exhibit a visible but highly
reddened AGN continuum and represent some of the most luminous known
AGN; see Section 2.1) a very high fraction ($\sim$80\%) are associated
with galaxy mergers and disturbances \citep[e.g.,][]{glik15qsomerge}.
Powerful, heavily-obscured {\em WISE}-selected quasars at $z\sim2$
\citep[e.g.,][]{asse15wiseqso} also
exhibit a large fraction of mergers \citep{fan17hotdog}. These results
are suggestive of a link between mergers and powerful obscured AGN, but
further work is needed to confirm this phenomenon.

In an evolutionary scenario, the same galaxy-scale dust and gas that
obscures the AGN may also be expected to
produce enhanced star formation. Any distinction between
the star-forming properties of obscured and unobscured AGN immediately rules out
the simplest unified AGN models, in which obscuration is purely due to
orientation of the dusty torus. Far-IR and submm studies of luminous quasars show that obscured sources exhibit stronger emission from cold dust; this conclusion holds for obscuration measured in X-rays \citep[e.g.,][]{page04submm, page11obsqso} and also from IR-optical SEDs \citep[e.g.,][]{chen15qsosf}. \citet{chen15qsosf} furthermore showed that the fraction of quasars that are obscured
increased strongly with far-IR luminosity ({\bf Figure \ref{fig:chen}}),
consistent with a picture in which obscuration in luminous AGN is frequently associated with
galaxy-scale dust. We emphasize, however, that these studies focused primarily on powerful quasars; for less luminous AGN classified as obscured or unboscured in the X-rays or optical, no comparable difference in average far-IR emission is observed \citep{merl14agnobs}. These results suggest that a connection between obscuration and galaxy-scale star-forming material may be most prevalent in powerful AGN.

A final piece of the evolutionary puzzle comes from spatial
correlation studies, which provide a robust statistical measure of the
large-scale structures (i.e., dark matter halos) in which galaxy and AGN
reside, independent of systematics in measurements of galaxy or AGN
properties \citep[e.g.,][]{berl02hod}. Differences in the host halo masses
between AGN types would rule out the simplest unified AGN models. 
Comparisons of obscured and unobscured AGN clustering have engendered
significant debate, with some studies showing stronger clustering for
obscured AGN, others for unobscured AGN, and still others showing no difference
\citep[e.g.,][]{hick11qsoclust, alle11xclust, mend16agnclust}. The large samples of $>10^5$ quasars identified with {\em
WISE} have enabled high-precision measurements, 
consistently showing stronger clustering for the obscured population
\citep[e.g.,][]{dono12wise, dipo14qsoclust}. This difference has been
confirmed through independent cross-correlations of the quasar positions
with lensing maps derived from the cosmic microwave background
\citep[e.g.][]{dipo15qsocmb,dipo17qsoclust}.  These results can be explained qualitatively
with a model in which the obscured quasars have
SMBHs that are undermassive relative to their halos and are ``catching
up'' to their final mass, consistent with an evolutionary scenario \citep{dipo17model}.

\section{IMPLICATIONS FOR OBSCURED AGN IN OBSERVATIONAL COSMOLOGY}

The previous section demonstrates that, while most AGN obscuration is likely to occur in nuclear regions within the sphere of influence of the SMBH (Section~\ref{sec:torus}), a significant fraction of the obscuration may originate on larger scales (Sections~\ref{sec:starburst} and \ref{sec:galaxy}). Obscuration during discrete events such as starbursts or galaxy mergers point to a link between SMBH growth and the cosmological formation of galaxies and large-scale structures. In this section we explore three of the implications of obscured SMBH growth for observational cosmology: (1) The SMBH-galaxy evolutionary sequence, (2) Obscured SMBH growth in the early Universe, and (3) The ``missing'' AGN population and the radiative efficiency of SMBH accretion.

\label{sec:cosmology}

\subsection{The evolutionary sequence and the SMBH-galaxy connection}

\label{sec:evolution}

\begin{figure}[t]
\includegraphics[width=\textwidth]{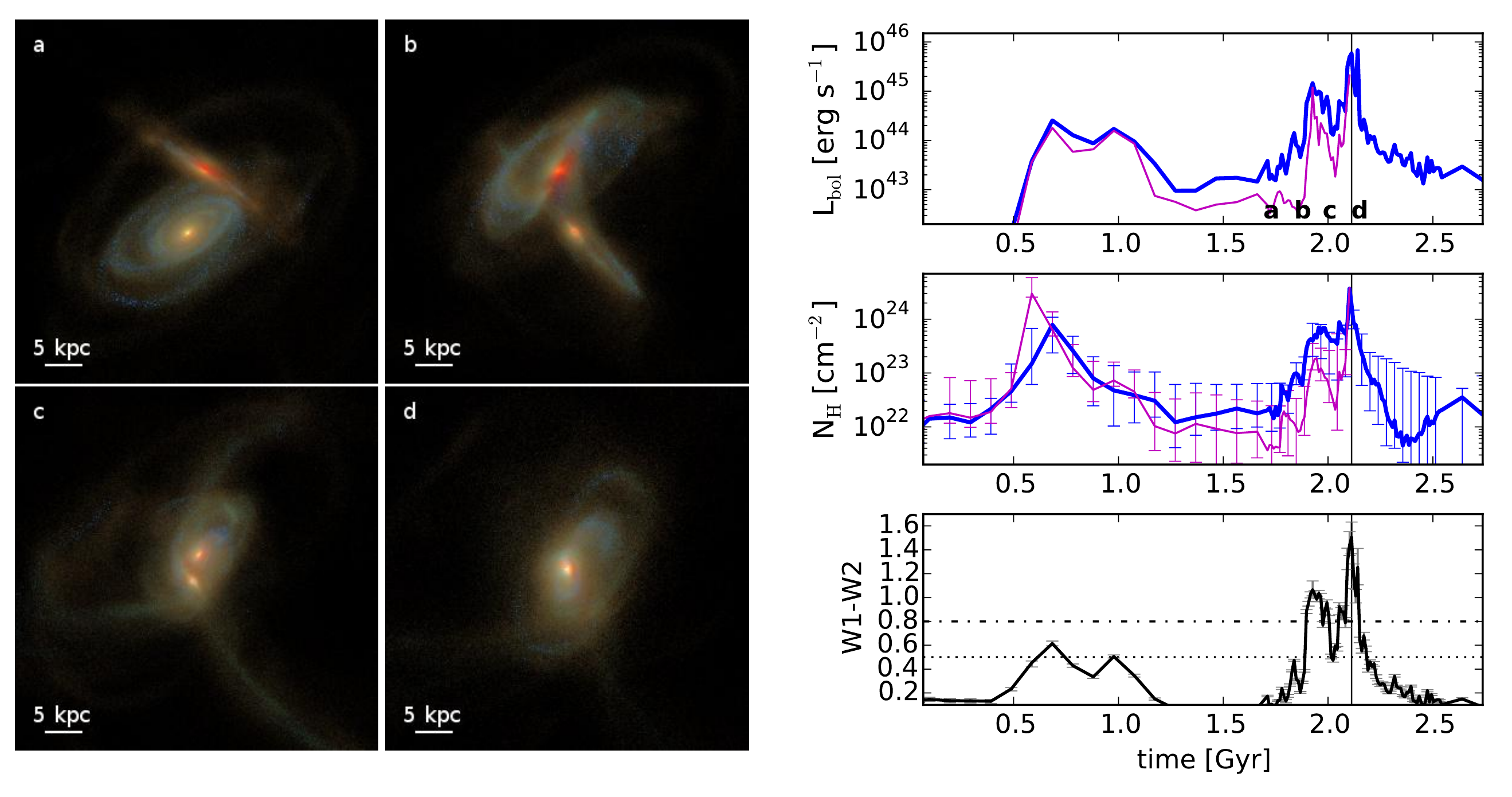}
\caption{An illustration of SMBH fueling and obscuration during a galaxy merger \citep{blec18merger}. The left panel shows images of simulated galaxies in four different stages of a merger, with the time dependence of AGN $L_{\rm bol}$, $N_{\rm H}$, and {\em WISE} W1--W2 color are shown in the right column. The coalescence of the galaxies and SMBHs produces a spike in SMBH accretion and $L_{\rm bol}$, along with a redder W1--W2 color and a jump in $N_{\rm H}$ that reaches approximately Compton-thick ($N_{\rm H}\sim10^{24}$ cm$^{-2}$) level. Figure from \citet{blec18merger}, courtesy of L.~Blecha. }
\label{fig:blecha}
\end{figure}

As discussed in Section \ref{sec:importance}, connections between SMBHs and galaxies in their cosmic evolution have attracted a great deal of interest, motivated in part by observations of concurrent AGN and starbursts in well-studied local systems \citep[e.g.,][]{sand96,farr03}, statistical connections between AGN activity and star formation or stellar mass in extragalactic surveys \citep[e.g.,][]{chen13agnsf, azad15primus}, and observed correlations between SMBH masses and host galaxy properties (velocity dispersion, stellar mass, etc.; \citealt{mcco13bhgal,korm13bh,grah16bhbulge}). The growth of SMBHs releases enormous amounts of energy in the form of radiation, outflows, and relativistic jets that can significantly influence the evolution of the host galaxies \citep[e.g.,][]{alex12bh, fabi12feed}. Many galaxy formation models require energy input from AGN to produce the observed population of quiescent galaxies \citep[e.g.,][]{bowe06gal, dubo16horizon}.  Ultimately, there is likely to be a complex interplay between SMBH and galaxy growth, in which AGN activity follows or enhances the growth of stars in some cases, and shuts off or prevents new star formation in others \citep[see][for a review]{harr17agnsf}. 

An important role in many models of SMBH-galaxy co-evolution is played by obscured AGN. Some models posit rapid phases of galaxy and SMBH growth triggered by mergers, interactions, or violent instabilities that can also disrupt the gas content of the galaxy and shroud the AGN \citep[e.g.,][]{sand88, hopk08frame1}. The majority of the SMBH and galaxy growth is predicted to occur in an early obscured phase (e.g., \citealt{blec18merger}; {\bf Figure~\ref{fig:blecha}}), followed by a ``blowout'' phase in which AGN feedback both limits SMBH growth and ejects gas from the galaxy potential, quenching the starburst and preventing further star formation \citep[e.g.,][]{ishi16agnsb}. In this picture the period of obscured AGN activity represents the key phase for building up the mass of the SMBH, while the subsequent AGN luminosity limits the growth and sets the relationship between the SMBH and galaxy. The impact of the AGN on surrounding gas can be observed in outflows, heating, and turbulence in molecular, atomic, and ionized gas \citep[e.g.,][]{gree11obsqso, harr14agnoutflow, feru15mrk231}. 

However, whether galaxy and SMBH evolution is primarily driven by discrete, dramatic phases of evolution remains unclear. Some studies have suggested that the relationships between SMBHs and galaxies progress slowly over cosmic time, with the fueling (and obscuration) of AGN being primarily a stochastic process \citep[e.g.,][]{cist11secular, mull12agnms}, so that obscured AGN do not represent an especially important phase in SMBH growth or feedback. It may also be possible that the importance of obscured AGN activity depends strongly on the type of galaxy and its evolutionary history; since massive ellipticals and galaxy bulges have old, $\alpha$-enhanced stellar populations that formed in rapid starbursts \citep[e.g.,][]{zhu10elliptical, mcder15atlas3d}, obscured AGN activity may be more important in the formation of these systems than in disk-dominated galaxies with more quiescent SF histories \citep[e.g.,][]{ishi17compact}. 

A key clue in uncovering the role of obscured AGN in the cosmological growth of SMBHs is the determination of whether obscuration is connected with processes in the nuclear torus (small enough to be decoupled from the broader galaxy formation) or on the scale of the galaxy.  Further observations are required to determine the fractions of AGN that are obscured due to material on ``torus'' and ``galaxy'' scales (as discussed in Section \ref{sec:galaxy}), and to determine the sub-populations of galaxies for which obscured AGN may play a particularly important role.

\subsection{The evolution of obscured SMBHs at high redshift}

\label{sec:highz}

\begin{figure}[t]
\includegraphics[width=\textwidth]{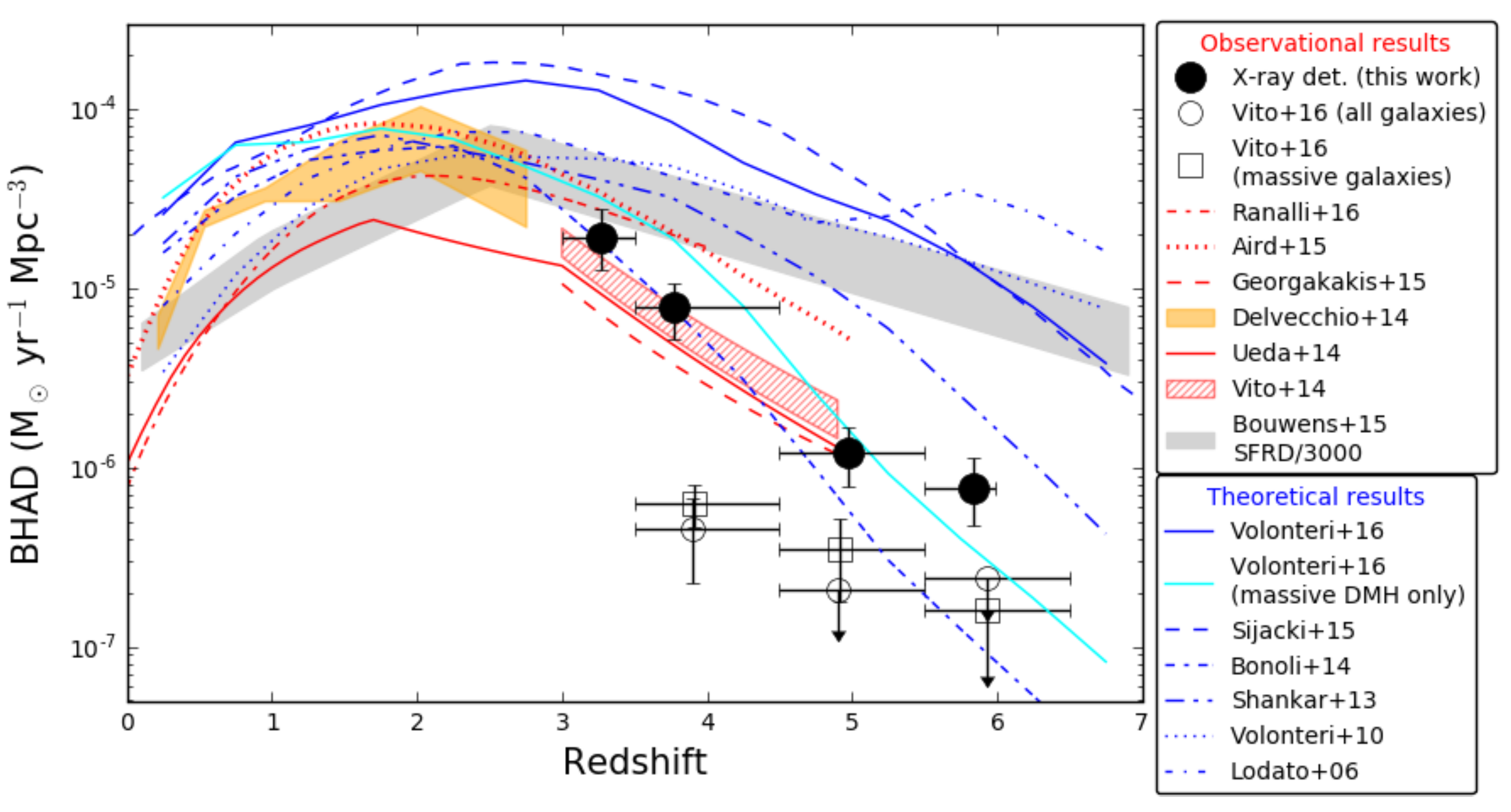}
\caption{Evolution in the AGN X-ray luminosity density at high $z$, as determined through {\em Chandra} Deep Field Observations \citep{vito18highz}. The rapid drop-off of the luminosity density (shown by the black points) suggests that SMBH growth falls with redshift faster than does star formation (shown by the gray filled region). This fast evolution rules out a number of SMBH evolution models (shown as blue lines); this tension may be resolved by the existence of a highly-obscured AGN population at these redshifts. Figure from \citet{vito18highz}, courtesy of F.~Vito. Observational and model results are from: \citet{vito16highz,rana16xlf,aird15xlf,geor15highz,vito14xlf,bouw15highz,volo16horizon,sija15illustris, bono14seed, shan13agn, volo10bhform, loda06seed} \label{fig:highz}.}
\end{figure}

If AGN obscuration is connected to the gas content of the host galaxy, then we may expect obscuration to be enhanced at higher redshift, where the fraction of mass of galaxies in atomic or molecular gas is far higher than in the local Universe \citep[e.g.,][]{cari13gas}.  As discussed in Section \ref{sec:demographics},  X-ray observations have shown hints that the obscured fraction may increase with redshift. For a galaxy with a high enough mass of gas, even large-scale ($\sim$kpc) obscuring clouds can be heavily absorbing or even Compton-thick, so X-ray observations may miss an increasing fraction of the SMBH growth at higher redshifts. Powerful, mid-IR-bright obscured quasars at $z\sim2$ frequently show weak or absent X-ray detections even in deep observations  \citep{ster14nustarwise, delm16qso}, suggesting that the Compton-thick fraction for these AGN are $\sim$25--50\%, comparable to local samples (see Section~\ref{sec:demographics}) but at higher AGN luminosities. If this high Compton-thick fraction is also present for lower luminosity AGN at high redshift, a substantial fraction of the SMBH growth at high redshift might not be captured by current X-ray observations (see Section~\ref{sec:cxb} for the implications of this on the global radiative efficiency).

The question of ``missing'' obscured AGN is particularly interesting at high redshift ($z>3$). Luminous quasars with $M_{\rm BH}>10^9$ $M_\odot$ are observed to emerge at $z=6$--7 \citep[e.g.,][]{mort11z7qso, wu15z6qso} and increase in space density to lower redshifts with a peak at $z\sim2$ \citep[e.g.,][]{kell10qsoedd}. The growth of lower-mass SMBHs that ultimately power these massive quasars can be probed with deep X-ray studies of lower-luminosity AGN \citep[see][for a review]{volo10bhform}. Recent studies of the deepest {\em Chandra} fields reveal a steep drop-off in the X-ray selected AGN space density at $z>3$, with the evolution strongest for soft X-ray luminosity $< 10^{44}$ erg s$^{-1}$ (\citealt{vito18highz}; see {\bf Figure \ref{fig:highz}}). If the X-ray observations do probe the complete radiative output of AGN at these redshifts, then models of SMBH ``seeds'' may encounter problems with insufficient SMBH growth to produce the observed massive quasars at lower redshifts. One potential solution is if the AGN are heavily obscured, which would allow for the presence of many more growing SMBHs that lie below the {\em Chandra} detection (or stacking) thresholds \citep[e.g.,][]{nova13bhacc, vito18highz}. 

At the highest redshifts, obscured accretion is an essential component to some of the SMBH seed models themselves. In particular ``direct collapse'' models posit the growth of early SMBHs in gas-rich dark matter halos  \citep[e.g.,][]{volo10quasistar, maye10collapse} and the accretion process in these models implies high covering factors with Compton-thick absorption. In these models, the earliest growth of SMBHs is necessarily heavily obscured, although some signatures of these direct-collapsing systems may be observable through reprocessed IR radiation \citep[e.g.,][]{nata17jwst}. 

\subsection{Obscured AGN, the cosmic X-ray background, and the radiative efficiency of black hole accretion}

\label{sec:cxb}

	\begin{figure}[t]

\includegraphics[width=3.6in]{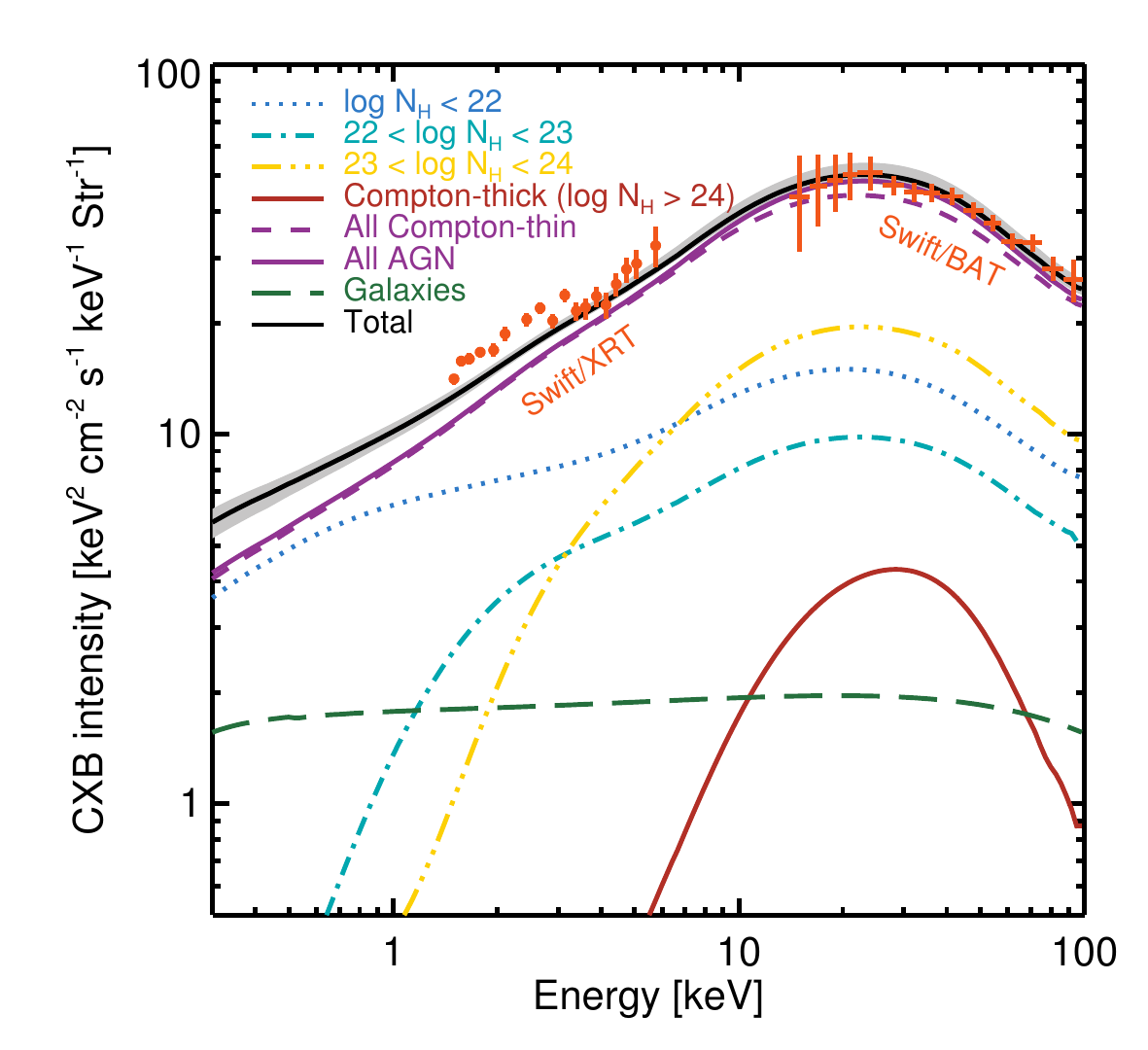}

\caption{Synthesis model of the cosmic X-ray background (CXB; \citealt{aird15xlf}), showing the contributions to the CXB from unobscured AGN (blue dotted line) and obscured AGN with varying levels of $N_{\rm H}$. The high-energy peak of the CXB is dominated by obscured sources, with a significant contribution from Compton-thick AGN (thick red line). CXB synthesis models have been used to estimate the SMBH radiative efficiency $\epsilon$, but may not account for the presence of an extremely obscured population (with $N_{\rm H}>10^{25}$ cm$^{-2}$) that does not contribute significantly to the CXB (Section~\ref{sec:cxb}).  Figure from \citet{aird15xlf}, courtesy of J.~Aird. }
\label{fig:cxb}
\end{figure}

The presence of a population of heavily obscured AGN has important consequences for the fundamental physics of SMBH accretion (in particular the radiative efficiency), and the corresponding origin of the cosmic background radiation. As first proposed by \cite{solt82}, the total radiation emitted by SMBHs over cosmic time provides a powerful clue to the accretion process that produces the population of SMBHs observed at low redshift. The concept is elegantly simple: The total radiation density produced by SMBHs ($U_{\rm T}$) is equal to the product of the mass density of SMBHs in the local Universe (assumed to have been accumulated via accretion) $\rho_{\rm SMBH}$ and the radiative efficiency $\epsilon$. Including a factor of $1-\epsilon$ to account for the mass-energy lost as radiation, this relationship can be written as:

\begin{equation}
\rho_{\rm SMBH} c^2 = U_{\rm T}\frac{1-\epsilon}{\epsilon}
\end{equation}

A number of early studies used the space density of optically-selected unobscured quasars to estimate $\epsilon$ \citep[e.g.,][]{yu02smbh}. These studies included no obscured AGN and so naturally produced on a lower limit on $U_{\rm T}$, or required an estimate of the fraction of radiation that was obscured, as well as an estimate of the optical bolometric correction (the scale factor $k_{\rm opt}$ to convert from the optical radiation density $U_{\rm opt}$ to the total radiation density; $U_{\rm T}=k_{\rm opt}U_{\rm opt}$). Subsequent studies attempted to directly account for the obscured sources through measurements of the cosmic X-ray background (CXB); sensitive X-ray observations have confirmed that the CXB is dominated by emission from individual AGN (e.g., \citealt{baue04, hick06a}, see \citealt{bran15xray} for a recent review) and dominate the CXB even to $E>10$ keV \citep[e.g.,][]{aird15nustar, harr16nustar}. Successful CXB synthesis models universally require a population of obscured AGN to produce the observed peak in the spectrum at $E\sim 30$ keV \citep[see {\bf Figure~\ref{fig:cxb}}]{gill07, trei09cxb, ball11nustar, akyl12cxb, ueda14cxb, aird15xlf}. After correcting the total CXB radiation for absorption (either empirically, or through AGN synthesis models), and assuming an X-ray bolometric correction, it can be used to estimate $\epsilon$ \citep[e.g.,][]{fabi99xrb}.

In recent years there has been substantial progress in understanding the cosmic synthesis of SMBHs. Large obscured AGN populations have been discovered that could substantially increase $U_T$ (Section \ref{sec:demographics}) and updates to the local SMBH mass density have come from new dynamical measurements of SMBH masses and re-assessments of relationships between SMBH masses and galaxy properties 
\citep[e.g.,][]{korm13bh}. These new scaling relations significantly increased the estimate of $\rho_{\rm SMBH}$, by up to a factor of 5; for the same value of $U_T$, this would lead to a corresponding decrease in $\epsilon$ that would fall uncomfortably far below the theoretically expected value of $\epsilon\approx0.1$ (\citealt{kerr63, shap83}; see Section~\ref{sec:intro}).   

However, in these analyses one fundamental uncertainty is the fraction of AGN that are so heavily obscured that they contribute little or nothing to the observed radiation fields used to compute $U_T$ and thus $\epsilon$ (see e.g., \citealt{mart09bh, nova13bhacc}). The total obscured fraction and $N_{\rm H}$ distribution is often inferred from local optical or X-ray studies \citep[e.g.,][]{burl11batagn, ricc17batagn}. However, in principle a substantial fraction of sources  may be missed through extremely heavy absorption that is not accounted for in the selection function (as discussed in Section \ref{sec:identification}), and the $N_{\rm H}$ distribution may evolve significantly with redshift. The ``missing'' AGN population could therefore contribute a substantially higher fraction of the total SMBH growth than is assumed based on local studies.  Due to energy conservation, the radiation from such sources must ultimately emerge at far-IR and submm wavelengths as it is reprocessed into thermal emission from cold dust. A census of the AGN population obtained from fitting broad-band SEDs (including far-IR data from {\em Herschel}) suggests that the evolution of the luminosity function of AGN identified in the IR is comparable to that determined from other wavebands \citep{delv14her}.  However, the signal from extremely heavily buried AGN may be challenging or impossible to distinguish from emission powered by star formation processes that completely dominate the IR background at $>10$ $\mu$m \citep{shi13bg}. 
\citet{coma15xrb} showed that the total radiation output from accreting SMBHs could be increased by a factor of $\sim$2 in the form of extremely Compton-thick, X-ray faint AGN without violating constraints in the X-ray and IR backgrounds. This, in turn, would increase $\epsilon$ by a similar factor, potentially resolving tension with theoretical expectations.

\section{CONCLUSIONS AND FUTURE PROSPECTS}

\label{sec:conclusions}

	In this review we have highlighted progress in identifying and characterizing obscured AGN, and have detailed the associated physical insights related to the accretion process onto SMBHs and its cosmological implications. Observationally, the combination of sensitive observations over a broad range of wavelengths (particularly the mid-IR and hard X-rays) have enabled a much improved census of the population of growing SMBHs in the past decade (Sections \ref{sec:identification} and \ref{sec:demographics}). Major recent advances include better understanding of the geometry and properties of the obscuring ``torus'' (Section \ref{sec:torus}), evidence for obscuration on host galaxy scales and connections to galaxy evolution (Sections \ref{sec:starburst}, \ref{sec:galaxy} and \ref{sec:evolution}),  direct measurements of the contribution of AGN to the hard CXB (Section \ref{sec:cxb}), and the discovery of a (possibly large) population of very heavily obscured AGN that may be particularly important at high redshift (Section \ref{sec:highz}). These results suggest an important role for obscured accretion in the growth histories of SMBHs (both in the rapid evolution of massive galaxies, and in the early growth of ``seed'' SMBHs), and indicate that while the bulk of obscuration appears to occur in compact regions within the sphere of influence of the SMBH, obscuration occurs over a wide range of scales and physical conditions. In this final section we discuss how our understanding of obscured AGN wil be further improved with future observational facilities and developments in theoretical models, and suggest some open questions that provide particularly promising opportunities for progress.

\subsection{Forecasts for future facilities}

\label{sec:facilities}

The coming decades will see an impressive array of new observational resources that will enhance our abilities to detect and characterize obscured AGN. Here we will discuss the prospects for a handful of upcoming or proposed facilities in each of the wavebands discussed in Section \ref{sec:identification}.

\subsubsection{UV--Near-IR} Spectroscopy provides the most widely-applicable
method for studying obscured AGN in the optical and NIR (Section
\ref{sec:optspec}). From the ground, new large-scale multi-object
spectrographs in the optical and near-IR such as DESI\fnm\fnt{http://desi.lbl.gov/} \citep{desi16}, 4MOST\fnm\fnt{https://www.4most.eu/} \citep{dejo144most}, and Subaru
PFS\fnm\fnt{http://pfs.ipmu.jp/} \citep{taka14pfs} will identify and characterize huge numbers of obscured AGN,
including rest-frame optical lines out to $z\sim3$--4. In addition,
new surveys with integral field units such as SAMI\fnm\fnt{https://sami-survey.org/} \citep{croo12sami}, CALIFA \citep{sanc12califa}\fnm\fnt{http://califa.caha.es}, and MaNGA\fnm\fnt{http://www.sdss.org/surveys/manga/} \citep{bund15manga} are enabling detailed
spatially resolved studies of NLR ionization and kinematics. From space, {\em JWST}\fnm \fnt{https://jwst.nasa.gov/} and
{\em WFIRST}\fnm\fnt{https://wfirst.gsfc.nasa.gov/} will provide extremely sensitive infrared
spectroscopy with high spatial resolution, allowing
characterization of the emission features in rest-frame optical (which
are critical for AGN diagnostics; Section \ref{sec:optical}) and also
host galaxy properties of obscured AGN out to high redshift. Further
into the future, the 30-m class telescopes such as GMT\fnm\fnt{https://www.gmto.org/}, TMT\fnm\fnt{http://www.tmt.org/}, and E-ELT\fnm\fnt{https://www.eso.org/sci/facilities/eelt/} will
perform high sensitivity observations of obscured AGN to high
redshifts, and the LUVOIR\fnm\fnt{https://asd.gsfc.nasa.gov/luvoir/} and HABEX\fnm\fnt{https://www.jpl.nasa.gov/habex/} concept space missions would carry
out sensitive optical spectroscopy and imaging of obscured AGN with
extremely high angular resolution. Finally, LSST\fnm\fnt{https://www.lsst.org/} will provide
extremely deep, wide-field optical imaging that, while not able to
efficiently select obscured AGN, will be powerful for the
characterization of AGN host galaxies and identifying rare changing-look
AGN (see Section \ref{sec:optspec}).

\subsubsection{X-rays} For statistical studies of obscured AGN, new generations
of X-ray telescopes will enable deep, wide-area surveys resulting in
very large samples of AGN. The {\em eROSITA} mission\fnm\fnt{http://www.mpe.mpg.de/eROSITA} will identify
millions of AGN over the whole sky \citep{merl12erosita}, including a
significant number of obscured sources (with follow-up spectroscopy
from 4MOST), although its ability to probe the complete obscured AGN
population will be limited by its energy response, which drops off more quickly at energies $>3$ keV when compared to {\em Chandra} and {\em XMM}. Further into the future, the {\em Athena} mission\fnm\fnt{http://www.the-athena-x-ray-observatory.eu/} \citep{barc15athena} will
enable sensitive X-ray imaging spectroscopy with sensitivity to
$\approx$12 keV, and over a wide 40' field of view. {\em Athena}
surveys will reach, over a large area, a confusion-limited source
detection flux limit approximately equal to that of the {\em Chandra}
Deep Fields \citep[e.g.,][]{xue11cdfs, luo17cdfs}, and will allow extremely sensitive spectroscopy of faint
and high-redshift AGN, providing a significant leap forward in the
selection of distant heavily obscured and CT AGN. The {\em Lynx} concept mission\fnm\fnt{https://wwwastro.msfc.nasa.gov/lynx/} (with higher spatial
resolution) would probe up to two orders of magnitude fainter with
imaging and spectroscopy of obscured AGN out to the highest redshifts
and providing constraints on SMBH seed models (see Section \ref{sec:highz})
and imaging of obscured AGN nuclei to better constrain the nature of
obscuring and reflecting material (see Section \ref{sec:torus}). Among a
number of proposed smaller X-ray missions, a particularly powerful
tool for studying obscured AGN is {\em HEX-P}\fnm\fnt{pcos.gsfc.nasa.gov/studies/rfi/Harrison-Fiona-RFI.pdf}, which would build on
the success of {\em NuSTAR} with higher-resolution imaging
 at $E>10$ keV.

For detailed studies of individual obscured AGN, the upcoming {\em XARM}
mission\fnm\fnt{https://heasarc.gsfc.nasa.gov/docs/xarm/} will provide calorimeter observations with exquisite energy
resolution to measure the strength and profile of the Fe K$\alpha$
line, Compton shoulder, and other features to constrain the obscuring
geometry. To best take advantage of {\em XARM}’s capabilities, simultaneous observations at harder ($>$10 keV) X-rays (for
example with {\em NuSTAR} if it is still operational) are required to provide the broad energy coverage for constraining the
strength and shape of the Compton-reflected continuum. Finally, we
will soon see the emergence of X-ray polarization measurements, first
with the {\em IXPE} mission\fnm\fnt{https://ixpe.msfc.nasa.gov/} and potentially on longer timescales with
{\em XIPE}\fnm\fnt{http://www.isdc.unige.ch/xipe/}. Polarization provides a unique capability for studying
scattering and reflection, and so can offer new constraints on the geometry
of the nuclear regions of obscured AGN \citep{mari18xpol}.

\subsubsection{Mid-IR} Dramatic breakthroughs in the mid-IR studies of obscured AGN will
come from {\em JWST}, which will greatly improve the sensitivity of
mid-IR spectroscopy (for characterizing optically faint or X-ray faint
AGN and identifying extremely weak or buried sources; see Section~\ref{sec:mirspec}) and photometry
(for identifying obscured AGN in faint, high-redshift galaxies; see Section~\ref{sec:mircont}). For
extremely high angular resolution observations, the new MATISSE
instrument\fnm\fnt{https://www.eso.org/sci/facilities/develop/instruments/matisse/.html} on the VLT Interferometer (with corresponding NIR
imaging from the GRAVITY instrument\fnm\fnt{https://www.eso.org/sci/facilities/paranal/instruments/gravity.html}) will enable superior imaging
capabilities for follow up imaging studies of the resolved
dust emission in nearby AGN (see Section \ref{sec:torus}).

\subsubsection{Far-IR--radio} ALMA\fnm\fnt{http://www.almaobservatory.org/} will continue to make leaps forward in our understanding of obscured AGN; in particular, the development of new submm line diagnostics (e.g., \citealt{aalt15hcn, iman16alma2}; See Section~\ref{sec:farir}) will enable the identification of excitation due to AGN in even the most heavily buried systems, particularly with anticipated improvements in sensitivity with future ALMA upgrades. ALMA will also be key in constraining spatially-resolved properties of the molecular torus such as the outer radius, gas mass, and kinematics, as discussed in Section~\ref{sec:torus}.  Sensitive AGN diagnostic studies can also be performed with the LMT\fnm\fnt{http://www.lmtgtm.org/} and CCAT\fnm\fnt{http://www.ccatobservatory.org/}, which reach similar depths to ALMA over wider fields of view, although with lower angular resolution.  From space, concept far-IR observatories such as {\em OST}\fnm\fnt{https://asd.gsfc.nasa.gov/firs/} and {\em SPICA}\fnm\fnt{http://www.spica-mission.org/} would provide sensitive, high-resolution observations of line emission to explore the connection between obscured AGN activity and star formation, in order to constrain obscuration on galaxy scales and test evolutionary models (Sections \ref{sec:galaxy} and \ref{sec:evolution}). The most exciting potential for future breakthroughs in the radio band comes from the SKA\fnm\fnt{https://skatelescope.org/} and its prescursor radio telescopes including LOFAR\fnm\fnt{http://www.lofar.org/}, ASKAP\fnm\fnt{http://www.atnf.csiro.au/projects/askap/}, MeerKAT\fnm\fnt{www.ska.ac.za/meerkat/}, MWA\fnm\fnt{http://www.mwatelescope.org/}, and HERA\fnm\fnt{http://reionization.org/}. These observatories are moving toward a dramatic improvement in the sensitivity of deep radio surveys and will eventually allow us to probe, even out to $z=5$--6 and beyond, faint radio luminosities for which non-jetted sources dominate the AGN number counts. In combination with deep optical and IR surveys, the SKA will enable vastly deeper and more powerful use of the ``radio-excess'' method, as well as (along with enhanced capabilities with VLBI) directly resolving compact bright compact cores. These will allow for the detection of AGN that are otherwise obscured or swamped by star formation processes (Section~\ref{sec:radio}).

\subsection{Prospects for theoretical models}

\label{sec:models}

Our understanding of obscured AGN will also see progress through advances in theoretical models, both on the scale of the accretion flow and the obscuring torus, and in the context of large-scale cosmological models of galaxy formation. 

On small scales, our understanding of AGN obscuration has been improved by advances in hydrodynamical simulations that capture the complex, dynamic, and multi-scale nature of the AGN central engine.  Better physical prescriptions for feedback \citep[e.g.,][]{hopk16agnfeed}, chemistry \citep[e.g.,][]{wada16torus}, and radiative transfer \citep[e.g.,][]{jud17disc} all yield a more complete picture of the physical origin and observational signatures of obscuring material in AGN.  Complementary to fully hydrodynamic models, there have also been improvements in phenomenological models of AGN tori (for example, including multi-phase gas consisting of both clumpy and smooth components; e.g., \citealt{sieb15torus, stal16torus}) and the associated radiative transfer calculations that provide input for interpreting the observed AGN SEDs. The future will see additional connections between hydrodynamic and phenomenological models. For example simulations can provide a framework for realistic ranges of torus opening angles, cloud sizes and optical depths that can then be used to create phenomenological models for constraining the properties of the observed AGN tori.

On the largest scales, we can make use of galaxy formation models, for which one direct and immediate application will be to better understand the strong selection effects involved in identifying obscured AGN. As illustrated in Sections \ref{sec:intro} and  \ref{sec:identification} and {\bf Figure \ref{fig:sedim}}, selection effects due to obscuration, host dilution, and physical changes in accretion present a complex multi-faced problem, and it can be challenging to reliably ``invert'' these effects from any given survey to recover the underlying cosmological AGN population. To address these issues, studies increasingly make use of forward modeling that begin with a known underlying galaxy and AGN population, and model their observational signatures in a range of wavebands. By adjusting the input parameters to match the observational data, it is possible to recover (with some assumptions) a reliable measure of the intrinsic AGN population. This approach has been used successfully in synthesis models that produce the CXB and the local SMBH mass function \citep[e.g.,][]{merl08agnsynth}, while also fitting the evolution of the observed XLF. 
However, in most cases, these models have assumed empirical parameters for the XLF without explicit connection between the AGN and their host galaxies and dark matter halos. Given the importance of host galaxy dilution on AGN selection (see Section \ref{sec:identification}), and the potentially critical role played by large-scale structures in driving galaxy-SMBH evolution (see Section \ref{sec:evolution}), more complete modeling of the AGN population in a cosmological context is warranted. Some studies have included SMBH accretion in cosmological hydrodynamic simulations \citep[e.g.,][]{mcal17eagle, wein17agn}, with some success modeling the observed distributions of SMBH mass and AGN luminosity. However, these models necessarily assume simplified sub-grid prescriptions for AGN accretion and feedback \citep[see the discussion in][]{negr17subgrid} and due to the computational expense, they are not able to explore a wide range of parameters for the underlying AGN population. Recently, models have added AGN to simulated galaxies that are drawn either from semi-analytic dark matter and galaxy formation simulations \citep[e.g.,][]{jone17agn}, or observed distributions in galaxy mass and star formation history \citep[e.g.,][]{weig17agn}. These prescriptions allow for the flexibility to include in galaxies a population of AGN with wide range of underlying parameters, while also modeling the full multiwavelength properties of the AGN including the emission and/or obscuration from the host galaxy. This modeling yields insights into which AGN are not selected in multiwavelength surveys, and provide useful tools for recovering the full population of obscured AGN.   

An ultimate goal for theoretical models of obscured AGN would be to perform a simulation over a dynamic range from the size of the galaxy or dark matter halo ($>$kpc) down to the accretion disk itself, covering all the relevant scales for gas dynamics and obscuration by gas and dust (Section \ref{sec:physics}). To date, computational limitations have meant that a full self-consistent treatment of the relevant physics over this range of scales has been out of reach. However, recent studies have performed 3D simulations of feeding, feedback, and obscuration from 0.1 pc to 100 pc scales \citep{hopk16agnfeed}, while  simulations of SMBH feeding and feedback can be now performed on a full galaxy scale, with resolution as small as 3 pc \citep[e.g.,][]{gabo13bhsim, negr17subgrid}, so a complete treatment across the full range of scales may be on the horizon.

\subsection{Outstanding questions}

\label{sec:questions}

This review has highlighted just a fraction of the exciting observational and theoretical progress that has greatly enhanced our understanding of obscured AGN in recent years. Looking to build on these insights, we conclude with some key outstanding questions:

\subsubsection{What is the physical origin of the torus?} The structural properties of the AGN dusty torus (such as a clumpy geometry, and broad range of covering factors, and the existence of dust along the poles) are coming into sharper focus (see Section \ref{sec:torus}). However, there remain many possible explanations for its formation, including warped accretion disks, AGN-driven winds, and starburst disks. Uncovering the physical origin (or multiple origins) of the torus is an important goal in coming years.

\subsubsection{What is the evolution of AGN obscuration and the connection to galaxy formation?} Observations have uncovered hints of evolution with redshift of the fraction of AGN that are obscured (see Section \ref{sec:demographics}), and suggest that some AGN are obscured by evolutionary processes in galaxies (see Section \ref{sec:evolution}) that, along with galaxy gas fractions and merger rates, can depend strongly with redshift. A major future objective will be to accurately measure the distribution of AGN obscuration with redshift, luminosity, or Eddington ratio, and to nail down the fraction of obscuration that is produced by galaxy-scale processes as opposed to a nuclear 
torus. One important clue may come from SMBH mergers identified by the {\em LISA} gravitational wave observatory\fnm\fnt{https://www.lisamission.org/}, which could place important constraints on galaxy and SMBH merger rates and provide signatures of obscured SMBHs that may not be identifiable from electromagnetic signatures. 

\subsubsection{How do we find the most heavily obscured AGN?}
Recent observations have begun to show evidence for a population of extremely heavily obscured AGN with $N_{\rm H}\gg10^{25}$ cm$^{-2}$. AGN at these levels of obscuration may exhibit few clear observational signatures (see Section \ref{sec:identification}) but might still contribute significantly to the global growth of SMBHs (see Section \ref{sec:cxb}). Some of the most promising techniques to uncover these sources use long-wavelength observations, for which the opacity is the lowest. For example, submm line diagnostics (see Section \ref{sec:farir}) or high-resolution detections of radio cores (see Section \ref{sec:radio}) may be critical in uncovering this most heavily obscured AGN population.

\subsubsection{What is the role of obscured accretion at the dawn of the first SMBHs?} 
The rapid early growth of SMBHs from ``seeds'' to massive quasar engines is currently an active area of inquiry. Theoretical models, as well as the rapid drop-off in the observed AGN density at high redshift, raise the possibility that a large fraction of the earliest SMBH growth may have been heavily obscured (see Section \ref{sec:highz}). Uncovering this population will be a major motivation for upcoming observatories that seek to understand the ultimate origin of SMBHs.

\section*{DISCLOSURE STATEMENT}
The authors are not aware of any affiliations, memberships, funding, or financial holdings that
might be perceived as affecting the objectivity of this review. 

\section*{ACKNOWLEDGMENTS}

R.C.H.\ acknowledges support from the National Science Foundation
through AST grant number 1515364 and CAREER grant
1554584, from NASA through grants NNX15AU32H and NNX16AN48G, and from
an Alfred P. Sloan Research Fellowship. D.M.A.\ acknowledges support
from the Science and Technology Facilities Council (STFC) for support
through grant ST/L00075X/1.  We have made extensive use of NASA's
Astrophysics Data System and the {\tt arXiv}. We are grateful to Cristina Ramos Almeida, Andrea Merloni, Michael DiPompeo, Lauranne Lanz, and Scientific Editor Luis Ho for valuable suggestions that improved the manuscript. Many of the ideas for the content of this
review were developed during the ``Hidden Monsters: Obscured AGN and
Connections to Galaxy Evolution'' workshop at Dartmouth College over
August 8--12 2016 and the ``Elusive AGN in the Next Era'' workshop at
George Mason University over June 12--15 2017. We would like to thank
the workshop participants for stimulating talks and discussions!


\input{ms.bbl}
\end{document}

%% file: aasjourn.tex
\newcommand\aj{{AJ}}%
\newcommand\araa{{ARA\&A}}%
\newcommand\apj{{ApJ}}%
\newcommand\apjl{{ApJ}}%
\newcommand\apjs{{ApJS}}%
\newcommand\aap{{A\&A}}%
\newcommand\aapr{{A\&A~Rev.}}%
\newcommand\mnras{{MNRAS}}%
\newcommand\pasp{{PASP}}%
\newcommand\pasj{{PASJ}}%
\newcommand\qjras{{QJRAS}}%
\newcommand\nat{{Nature}}%
\newcommand\nar{{New Astron. Revs}}%